%% file: mnras_template.tex
\documentclass[fleqn,usenatbib]{mnras}

\usepackage{newtxtext,newtxmath}

\usepackage[T1]{fontenc}

\DeclareRobustCommand{\VAN}[3]{#2}
\let\VANthebibliography\thebibliography
\def\thebibliography{\DeclareRobustCommand{\VAN}[3]{##3}\VANthebibliography}

\usepackage{tabularx}
\usepackage{newtxtext,newtxmath}
\usepackage{graphicx}	% Including figure fi
\usepackage{amsmath}	% Advanced maths commands
\usepackage{multirow}
\usepackage{subcaption}

\input{include_tex/definitions}

\input{include_tex/additional_def}
\usepackage{xcolor}
\defcitealias{Barai2018}{B18}
\defcitealias{Valentini:2021}{V21}
\title[Probing $z \gtrsim 6$ massive black holes with gravitational waves]{Probing $z \gtrsim 6$ massive black holes with gravitational waves}

\author[Chakraborty et al.]{
Srija Chakraborty,$^{1}$\thanks{E-mail: srija.chakraborty@sns.it}
Simona Gallerani,$^{1}$
Tommaso Zana,$^{1}$ 
Alberto Sesana$^{2,3}$
\newauthor{Milena Valentini,$^{4,5,6,7}$}
David Izquierdo-Villalba,$^{2,3}$
Fabio Di Mascia,$^{1}$
Fabio Vito $^{8}$
\newauthor{and Paramita Barai $^{9}$}
\\
$^{1}$Scuola Normale Superiore, Piazza dei Cavalieri 7, 56126 Pisa PI, Italy\\
$^{2}$Dipartimento di Fisica “G. Occhialini”, Università degli Studi di Milano-Bicocca, Piazza della Scienza 3, I-20126 Milano, Italy\\
$^{3}$INFN, Sezione di Milano-Bicocca, Piazza della Scienza 3, 20126 Milano, Italy\\
$^{4}$ Universit{\"a}ts-Sternwarte, Fakult{\"a}t f{\"u}r Physik,  Ludwig-Maximilians Universit{\"a}t  M{\"u}nchen, Scheinerstr. 1, D-81679 M{\"u}nchen, Germany\\
$^{5}$ Excellence Cluster ORIGINS, Boltzmannstr. 2, D-85748 Garching, Germany\\
$^{6}$ INAF - Osservatorio Astronomico di Trieste, via Tiepolo 11, I-34131 Trieste, Italy\\
$^{7}$Astronomy Unit, Department of Physics, University of Trieste, via Tiepolo 11, I-34131 Trieste, Italy\\
$^{8}$ INAF - Osservatorio Astronomico di Bologna, via Piero Gobetti 93/3, I-40129 Bologna, Italy\\
$^{9}$ Centro de Ci\^{e}ncias Naturais e Humanas - Universidade Federal do ABC (CCNH-UFABC), Av. dos Estados 5001, Santo Andr\'{e} - SP, 09210-580, Brazil\\
}

\date{Accepted XXX. Received YYY; in original form ZZZ}

\pubyear{2022}
\DeclareUnicodeCharacter{2212}{-}
\begin{document}
\label{firstpage}
\pagerange{\pageref{firstpage}--\pageref{lastpage}}
\maketitle

\begin{abstract}
\\We investigate the coalescence of massive black hole ($M_{\rm BH}\gtrsim 10^{6}~\rm M_{\odot}$) binaries (MBHBs) at $6<z<10$ by adopting a suite of cosmological hydrodynamical simulations of galaxy formation, zoomed-in on biased ($ >3 \sigma$) overdense regions ($M_h\sim 10^{12}~\msun$ dark matter halos at $z = 6$) of the Universe. We first analyse the impact of different resolutions and AGN feedback prescriptions on the merger rate, assuming instantaneous mergers. Then, we compute the halo bias correction factor due to the overdense simulated region. Our simulations predict merger rates that range between 3 -- 15 $\rm yr^{-1}$ at $z\sim 6$, depending on the run considered, and after correcting for a bias factor of $\sim 20-30$.

For our fiducial model, we further consider the effect of delay in the MBHB coalescence due to dynamical friction. We find that 83 per cent of MBHBs will merge within the Hubble time, and 21 per cent within 1 Gyr, namely the age of the Universe at $z > 6$.
We finally compute the expected properties of the gravitational wave (GW) signals and find the fraction of LISA detectable events with high signal-to-noise ratio (SNR $>$ 5) to range between 66-69 per cent. However, identifying the electro-magnetic counterpart of these events remains challenging due to the poor LISA sky localization that, for the loudest signals ($\mathcal M_c\sim 10^6~\msun$ at $z=6$), is around 10~$\rm deg^2$.

\end{abstract}


%%%%%%%%%%%%%%%%%%%%%%%%%%%%%%%%%%%%%%%%%%%%%%%%%%

%%%%%%%%%%%%%%%%% BODY OF PAPER %%%%%%%%%%%%%%%%%%
\hbadness=99999
\section{Introduction}
\label{sec:Introduction}

It is now widely agreed that that the centres of nearby galaxies host massive black holes (MBHs, $10^{6}\lesssim M_{\rm BH}\lesssim 10^{10}~\rm M_{\odot}$) whose masses correlates with several properties of the host galaxy itself \citep[e.g.][]{Magorrian_1998,Kormendy_2001}. Observational evidences also indicate the existence of bright quasars at $z\gtrsim 6$ (e.g. \citealt{Fan_2006,Jiang_2010}) powered by super massive black holes (SMBHs, $10^{8}\lesssim M_{\rm BH}\lesssim 10^{10}~\rm M_{\odot}$) that are accreting close to the Eddington rate. It is still a theoretical challenge to explain how such SMBHs have been assembled within 1 Gyr after the birth of the Universe.

These bright, high-$z$ quasars are expected to be located in massive dark matter (DM) halos ($M_{\rm halo}\gtrsim 10^{12}~\rm M_{\odot}$). According to the hierarchical structure formation scenario (e.g. \citealt{White_Rees,Peebles,Blumenthal:1984bp}), such massive DM halos result from the mergers of smaller halos that formerly harboured the initial seeds of the SMBHs we observe at $z\sim 6$. 
Several candidates have been proposed so far as SMBH seeds: (i) {\it light seeds} ($M_{\rm seed}\sim 10-100~\rm M_{\odot}$), formed as remnants of Pop III stars (\citealt{Madau1998ApJ...498..106M,Heger_2003,Yoshida_2008,Hirano_2015}) 
at $z\sim 20-30$; (ii) {\it intermediate seeds} ($M_{\rm seed}\sim 1000~\rm M_{\odot}$), produced in compact nuclear star clusters as a consequence of runaway stellar mergers at $z\sim 10-20$ (\citealt{Davies2011ApJ...740L..42D,Devecchi2012MNRAS.421.1465D,Lupi_2014,Mapelli2016MNRAS.459.3432M,Reinoso2018A&A...614A..14R}); (iii) {\it heavy seeds} or {\it direct collapse black holes} (DCBHs, $M_{\rm seed}\sim 10^4-10^6~\rm M_{\odot}$), resulting from the rapid collapse of metal poor gas clouds in $z\gtrsim 10$ atomic cooling halos (virial temperature $T_{\rm vir}\geq 10^4~\rm K$) where star formation is prevented by intense H$_2$ photo-dissociating Lyman Werner (LW) radiation (\citealt{Haehnelt_1994,Loeb_1994,Eisenstein_1995,Silk:1997xw,Shang2010MNRAS.402.1249S,Johnson_2012,Yue_2013,Ferrara_2014}).

Numerous uncertainties remain associated to each of the scenarios discussed above (e.g.\citealt{Volonteri_2003,Koushiappas_2004,Begelman_2006,Lodato_2006,Volonteri_2012,Tanaka_2009,Latif_2013,Woods_2019}). To understand which of the aforementioned scenarios is the most promising in order to explain the presence of SMBHs at $z\sim 6$ it is necessary to study the still undiscovered population of Intermediate-Mass black holes (IMBHs, $10^{5}\lesssim M_{\rm BH}\lesssim 10^{7}~\rm M_{\odot}$) at $6\lesssim z \lesssim 10$. Various works have investigated the possibility that these IMBHs lie at the centers of dwarf galaxies (\citealt{Reines2013,Silk_2017,Barai_2019}), and it is difficult to detect them at high-z because of low surface brightness of the galaxies as well as the low intensity of the IMBHs.

The proposed space-borne GW observatory Laser Interferometer Space Antennae (LISA; \citealt{consortium2013gravitational}), due to be launched in 2034, is potentially an exquisite tool to accomplish this goal. LISA is in fact designed to be sensitive to signals in the frequency range $10^{-4} - 10^{-1}$ Hz, thus capable to detect GWs from massive ($10^4-10^8~\rm M_{\odot}$) BH binaries at very-high redshift \citep[even $z\sim 20$ if BHs in this mass range already exist at such early epochs;][]{ Haehnelt_1994,Jaffe_2003,Wyithe_2003,Sesana_2004,Sesana_2005,2022arXiv220306016A}.
Different scenarios of BH seeds are expected to leave signatures in all those observables that can be probed by GW observations of MBHBs \citep[e.g. merger rates, BH mass distribution;][]{Bhowmick}. 

Several studies have used semi-analytical models to estimate the MBHB merger rate.
For instance, \citet[]{Sesana_2007} made predictions for the merger rate of MBHs when the compact objects start their evolution either as DCBHs or as Pop III star remnants.
In this study, the authors considered different stellar feedback regimes to trace the metal enrichment in the gas halos of MBH formation.
Differently, \citet[]{Hartwig_2018} quantified the rate of in-situ mergers from binaries of DCBHs.
\citet{Dayal2019} investigated the dependence of the merger rate on the seed mass (light versus heavy seeds), the merger prescription (instantaneous versus delayed merger), and cosmic reionization. 
\citet{Barausse_2020} focussed on different seeding models, including various astrophysical processes, such as supernovae feedback. Comparably, numerous semi-analytical models, such as %Volonteri_2003,Koushiappas_2004,Begelman_2006,
\citet{Begelman_2006, Arun_2009, Klein_2016,Bonetti,Valiante_2020}, explored different evolutionary channels of MBHB formation (seed mass, accretion efficiency, metal enrichment, merger timescales) that could be detected by milli-Hz GW facilities, like LISA.

A different approach instead takes advantage of cosmological hydrodynamical simulations \citep[e.g.][]{Di_Matteo_2012,Vogelsberger2014Natur,Schaye2015MNRAS, Volonteri_2017} to study the dynamics of gas and its effect on the evolution of high redshift quasars.
\citet[]{Salcido} studied the SMBH mergers occurring in a fully cosmological simulation.
In particular, the authors looked for the expected LISA detection rate in the EAGLE run \citep[][]{Schaye2015MNRAS}, including also a prescription to delay the mergers.
More recently, several works made predictions for the LISA detection rate taking advantage of the \texttt{Illustris} simulations \citep{nelson2015}. Particularly, \citet[]{Katz_2019} analysed the effect of different evolutionary models for MBHB and studied the detectability of the merger events \citep[see also][]{katz2019_detectability}.
\citet{DeGraf2020} studied how different seeding models can affect the statistical properties of the MBH population, and the resulting merger rate. Furthermore, \citet{DeGraf2021} explored the relation between SMBH mergers and the morphology of their host galaxies at $z\leq 4$.

In this work, we adopt cosmological zoom-in hydrodynamical simulations of galaxy formation, based on the GADGET-3 code, to investigate the coalescence of massive black hole ($M_{\rm BH}\gtrsim 10^{6}~\rm M_{\odot}$) binaries at $6<z<10$. We consider a suite of simulations which differ both in terms of resolution and in the stellar/AGN feedback prescriptions implemented and investigate the impact of resolution and feedback on the merger rates of massive black hole binaries. For our fiducial model, we further investigate the associated GW properties (chirp mass, merger rate, characteristic strain, signal-to-noise ratio, angular resolution) of the MBHBs. We also quantify the effect of considering delays in MBHB merging due to dynamical friction on such GW properties. The paper is organized as follows:
in Section~\ref{sec:simulations}, we describe the numerical simulations analysed in this work; in Section~\ref{sec:gw}, we study the merger rate resulting from our simulations and in Section~\ref{sec:delay} we analyse the effect of adding delays in the MBHB coalescence in post-processing. In
Section~\ref{sec:GWhighz} we analyze the GW properties from the MBHB mergers.
Finally, we discuss our results in Section~\ref{sec:discussion} and draw our conclusions in Section~\ref{sec:conclusions}.

\section{Simulations}
\label{sec:simulations}
In this section, we describe the cosmological hydrodynamical simulations adopted in this work. We select simulations from the suites introduced by \citet{Valentini:2021} -- hereafter \citetalias{Valentini:2021} -- and \citet{Barai2018} -- hereafter \citetalias{Barai2018}.

The simulations are performed with the TreePM (particle mesh) + SPH (Smoothed Particles Hydrodynamics) code \code{GADGET-3}, an evolution of the public \code{GADGET-2} code \citep{Springel2005Gadget}, 
and follow the evolution of a $\sim 10^{12}$~M$_{\odot}$ halo at $z= 6$.
%featuring the improved SPH formulation by \citet{Beck2016MNRAS.455.2110B}.
In particular, we consider the following runs from \citetalias{Valentini:2021}: 
\begin{enumerate}
    \item[$\bullet$] \AGNfiducial{}: our fiducial model, featuring thermal AGN feedback;
    \item[$\bullet$] \BHsnoFB{}: a control run, analogous to AGN\_fid,  in which BHs and their accretion are included, but AGN feedback is turned off;
\end{enumerate}
and the following runs from \citetalias{Barai2018}:
\begin{enumerate}
     \item[$\bullet$] \AGNcone{}: in which the kinetic feedback is distributed in a bi-cone with and half-opening angle of ${45}$~\textdegree.
    \item[$\bullet$] \AGNsphere{}: featuring isotropic, kinetic AGN feedback;
\end{enumerate}

We summarise in the following sections the other main features of the simulations that are relevant for the present study, while we refer to the aforementioned papers for details.

\subsection{Valentini et al. 2021 models: \AGNfiducial{} and \BHsnoFB{}}
\label{sec:Valentini_sims}
\subsubsection{Initial conditions and resolution}
\label{subsec:Valentini_ICs}
The initial conditions are generated with the code \code{MUSIC}\footnote{\code{MUSIC}–Multiscale Initial Conditions for Cosmological Simulations: \url{https://bitbucket.org/ohahn/music}.} \citep{Hahn_2011}, assuming a $\Lambda$CDM cosmology\footnote{We adopt the following parameters by the \citet{Planck2016}: ${\Omega_{\rm M,0}= 0.3089}$, ${\Omega_{\rm \Lambda,0}= 0.6911}$, ${\Omega_{\rm B,0}= 0.0486}$, ${H_0 = 67.74~\rm{km~s}^{-1}~{\rm Mpc}^{-1}}$.}. First, a dark matter (DM)-only simulation is run from $z= 100$ to $z= 6$, with DM particles having a mass of $9.4\times 10^8~\msun$ in a comoving volume of $(148~{\rm Mpc})^3$. Then, a halo as massive as $M_{\rm halo}=1.12 \times 10^{12}~\msun$ at $z= 6$ is selected for a zoom-in procedure, to run the full hydrodynamical simulation. In the zoom-in region, the highest resolution particles have a mass of $m_{\rm DM}=1.55\times 10^6~\msun$ and $m_{\rm gas}=2.89 \times 10^5~\msun$. The gravitational softening lengths are\footnote{We use the following convention when indicating distances: a letter \emph{c} before the corresponding unit refers to \emph{comoving} distances (e.g. ckpc), while the letter \emph{p} refers to \emph{physical} units (e.g. pkpc). When not explicitly stated, we are referring to physical distances.} $\epsilon_{\rm DM}=0.72$~ckpc and $\epsilon_{\rm bar}=0.41$~ckpc for DM and baryon particles, respectively.
\begin{table*}
\caption{Summary of the main features of the suites of the cosmological hydrodynamical simulations adopted in this work (\citealt{Valentini:2021},\citealt{Barai2018})}
\label{simuls}
\centering
\resizebox{\textwidth}{!}{%
\begin{tabular}{|c|cc|cc|}
\hline
\textit{} & \multicolumn{1}{c|}{\textbf{\AGNfiducial{}}} & \textbf{\BHsnoFB{}} & \multicolumn{1}{c|}{\textbf{\AGNcone{}}} & \textbf{\AGNsphere{}} \\ \cline{2-5} 
 z=6 & \multicolumn{2}{c|}{\begin{tabular}[c]{@{}c@{}}Mass resolution[$\rm M_{\odot}$]:\\ $ m_{\rm DM}=1.5\times 10^6$\\ $ m_{\rm gas}=2.9\times 10^5$\\ Gas particle smoothing length[pc]=59\\ Size of the zoomed region=5.25cMpc\\ DM halo host:\\ $ M_{\rm halo}=1.1\times 10^{12}\rm M_{\odot}$\\ $\epsilon_r$ = 0.03 \\ \citetalias{Valentini:2021}\end{tabular}} & \multicolumn{2}{c|}{\begin{tabular}[c]{@{}c@{}}Mass resolution[$\rm M_{\odot}$]:\\ $m_{\rm DM}=7.54\times 10^6$\\ $ m_{\rm gas}=1.41\times 10^6$\\ Gas particle smoothing length[pc]=211\\ Size of the zoomed region=5.21cMpc\\ DM halo host:\\ $ M_{\rm halo}=4.4\times 10^{12}\rm M_{\odot}$\\ $\epsilon_r$ = 0.1 \\ \citetalias{Barai2018}\end{tabular}} \\ \hline
$M_{\star}$[$\msun$] & \multicolumn{1}{c|}{$4 \times 10^{10}$} & $3.5 \times 10^{10}$ & \multicolumn{1}{c|}{$7 \times 10^{10}$} & $6 \times 10^{10}$ \\ \hline
\begin{tabular}[c]{@{}c@{}}SFR [$\msun\rm yr^{-1}$]\end{tabular} & \multicolumn{1}{c|}{200} & 190 & \multicolumn{1}{c|}{200} & 300 \\ \hline
\begin{tabular}[c]{@{}c@{}}$ M_{\rm BH}[\msun]$\end{tabular} & \multicolumn{1}{c|}{$ 10^{9}$} & $5 \times 10^{11}$ & \multicolumn{1}{c|}{$2 \times 10^{9}$} & $5 \times 10^{8}$ \\ \hline
\begin{tabular}[c]{@{}c@{}}BHAR[$\msun\rm yr^{-1}$]\end{tabular} & \multicolumn{1}{c|}{35} & $3 \times 10^{4}$ & \multicolumn{1}{c|}{89} & 3 \\ \hline
Feedback & \multicolumn{1}{c|}{Stellar, AGN(thermal)} & Stellar & \multicolumn{1}{c|}{\begin{tabular}[c]{@{}c@{}}Stellar, AGN(kinetic,\\ bi-conical geometry)\end{tabular}} & \begin{tabular}[c]{@{}c@{}}Stellar, AGN(kinetic,\\ spherical geometry)\end{tabular} \\ \hline
\end{tabular}%
}
\end{table*}
\subsubsection{Sub-resolution physics}
\begin{enumerate}
\item[$\bullet$]\ \textbf{Cooling, star formation and stellar feedback{}}: the multiphase interstellar medium (ISM) is described by means of the MUlti Phase Particle Integrator (MUPPI) sub-resolution model \citep{Murante2010, Murante2015, Valentini2017, Valentini2019}. It features metal lines cooling, an \HH-based star formation, thermal and kinetic stellar feedback, the presence of an UV background, and the \citet{Tornatore_2007} model for chemical evolution. 

\item[$\bullet$]\ \textbf{Black holes seeding and merging{}}: BHs are treated as collisionless sink particles. Seeds of mass $M_{\rm BH, seed} = 1.48 \times 10^5~\msun$ are implanted in DM halos with mass exceeding $M_{\rm DM, seed}=1.48 \times 10^9~\msun$. This seeding prescription is meant to mimic in a simplistic way the DCBH scenario described in the \nameref{sec:Introduction} section. Two BHs are allowed to merge when their relative distance becomes smaller than twice the BH gravitational softening length, and their relative velocity is lower than the sound speed of the local ISM. The final BH is set on the position of the most massive BH which underwent the merger.
BH repositioning (or \emph{pinning}) is implemented, in order to prevent BHs from wandering from the centre of the halo in which they reside: at each time-step BHs are shifted towards the position of minimum gravitational potential within their softening length \citep[as also done in e.g.][]{Booth_2009, Schaye2015MNRAS, Weinberger2017, Pillepich2018}.

\item[$\bullet$]\ \textbf{Gas accretion on BHs{}}: besides BH-BH mergers, black holes are also allowed to grow via gas accretion, as described by the classical Bondi-Hoyle-Lyttleton (BHL) model \citep{Hoyle1939, Bondi_Hoyle, Bondi52, Edgar_2004}:
\begin{equation}  
\label{eq-Mdot-Bondi} 
\dot{M}_{\rm Bondi} = \frac{4 \pi G^2 M_{\rm BH}^2 \rho}{ \left(c_{\rm s}^2 + v^2\right) ^ {3/2}} , 
\end{equation}
where $G$ is the gravitational constant, $M_{\rm BH}$ is the BH mass, $\rho$ is the gas density, $c_{\rm s}$ is the sound speed, and $v$ is the velocity of the BH relative to the gas. These quantities are evaluated by averaging over the SPH gas particles within the BH smoothing length, with kernel-weighted contributions. 
Eq. \ref{eq-Mdot-Bondi} is used to estimate the contribution to the accretion rate from the cold and hot phase of the ISM, separately \citep{Steinborn2015, Valentini2020}. Accretion from the cold gas is reduced by taking into account its angular momentum \citep[see][for details]{Valentini2020}. 
The BH accretion rate is capped to the Eddington accretion rate.

\item[$\bullet$]\ \textbf{Quasar feedback{}}: a fraction of the accreted rest-mass energy is radiated away with a radiative efficiency $\epsilon_{\rm r}$, thereby providing a bolometric luminosity for an accreting BH equals to:
\begin{equation} 
\label{eq-Lr-BH} 
L_{\rm bol} = \epsilon_{\rm r} \dot{M}_{\rm BH}c^2,
\end{equation}
where $c$ is the speed of light and $\epsilon_{\rm r} = 0.03$ (\citealt{Sadowski2017}). Then, a fraction $\epsilon_{\rm f} = 10^{-4}$ (\citetalias{Valentini:2021}) of the radiated luminosity $L_{\rm bol}$ is coupled thermally and isotropically to the gas surrounding the BH. The AGN feedback energy is distributed to the hot and cold phases of the multiphase gas particles within the BH smoothing volume \citep{Valentini2020}.
\end{enumerate}

\subsection{Barai et al. 2018 simulations: \AGNcone{} and \AGNsphere{}} %\SG{elsewhere in italic! keep a coherent notation.}}
\label{sec:Barai_sims}
\subsubsection{Initial conditions and resolution}
Initial conditions are generated as in \citetalias{Valentini:2021} with the code \code{MUSIC} and adopting the same cosmology. The parent, DM-only simulation follows a comoving volume of $(500~{\rm Mpc})^3$, with DM particles having $m_{\rm DM}=2 \times 10^{10}~\msun$. The zoom-in run focuses on a DM halo as massive as $M_{\rm halo}=4.4 \times 10^{12}~\msun$; the highest resolution particles have $m_{\rm DM}=7.54 \times 10^6~\msun$ and $m_{\rm gas}=1.41\times 10^6~\msun$, with a gravitational softening length $\epsilon_{\rm bar}= \epsilon_{\rm DM}=1.48$~ckpc.
\subsubsection{Sub-resolution physics}
\begin{enumerate}
\item[$\bullet$]\ \textbf{Cooling, star formation and stellar feedback{}}: radiative heating and cooling is accounted for by employing the CLOUDY cooling tables computed by \citet{Wiersma2009MNRAS}. Star formation is implemented following the ISM multiphase model by \citet{Springel_2003}, in which an hot and a cold phase co-exist in pressure equilibrium, and assuming a density threshold for the star formation of $n_{\rm SF} = 0.13~\cc$. A \citet{Chabrier_2003} initial mass function (IMF) in the mass range $(0.1-100)~\msun$ is adopted. Stellar evolution and chemical enrichment are computed following \citet{Tornatore_2007}.%\SG{check all references}
\item[$\bullet$]\ \textbf{Black hole seeding and merging{}}: 
As in \citetalias{Barai2018}, the theoretical mass of seed BHs is $10^5 M_{\odot}$. However, their dynamical mass is much smaller in the \citetalias{Valentini:2021} simulations ($\sim 10^5~\rm m_{\odot}$ in \citetalias{Valentini:2021} versus $\sim 10^7 \msun$ in \citetalias{Barai2018}). The prescription is in fact more refined in \citetalias{Valentini:2021}, where seeded BHs are linked to stellar particles instead of DM particles. We further discuss this point in Sec. \ref{sec:mrates}.
\item[$\bullet$]\ \textbf{Gas accretion on BHs{}}: as in \citetalias{Valentini:2021}, the BHL model is adopted. However, the lower resolution of the \citetalias{Barai2018} simulations does not allow to properly describe the accretion process: thus Eq. \ref{eq-Mdot-Bondi} is multiplied by a numerical boost factor $\alpha=100$ \citep{Springel2005Modelling, Sijacki2009, Vogelsberger2014Natur}. No angular momentum effects are included in the \citetalias{Barai2018} formalism, and no distinction between the hot and cold gas phases is considered.
\item[$\bullet$]\ \textbf{Quasar feedback{}}: black holes are assumed to radiate energy away with an efficiency of $\epsilon_{\rm r}=0.1$, and a fraction $\epsilon_{\rm f} = 0.05$ of this energy is coupled to the surrounding gas via kinetic feedback as an energy-driven wind (see \citet{Barai2018} for details). The geometry of the feedback is bi-conical (i.e. energy is injected onto a bi-cone with a half-opening angle of ${45}$~\textdegree) in \AGNcone{} and spherical (i.e. energy distributed isotropically) in \AGNsphere{}. BHs grow $\sim 10$ times more massive at $z= 6$ in the \AGNcone{} case than in the \AGNsphere{} run as shown in top panels of fig: 2 in \citetalias{Barai2018}. This is because in the \AGNcone{} run more gas can inflow along the perpendicular direction to the bi-cone, and accrete onto the black hole.
\end{enumerate}

A summary of the two different simulation models are provided in Table \ref{simuls} for a comprehensive view. 
\section{Merger rate}
\label{sec:gw}
\subsection{Merger rate from overdense regions}\label{sec:mrates}
\begin{figure}
	
	\includegraphics[width=\columnwidth]{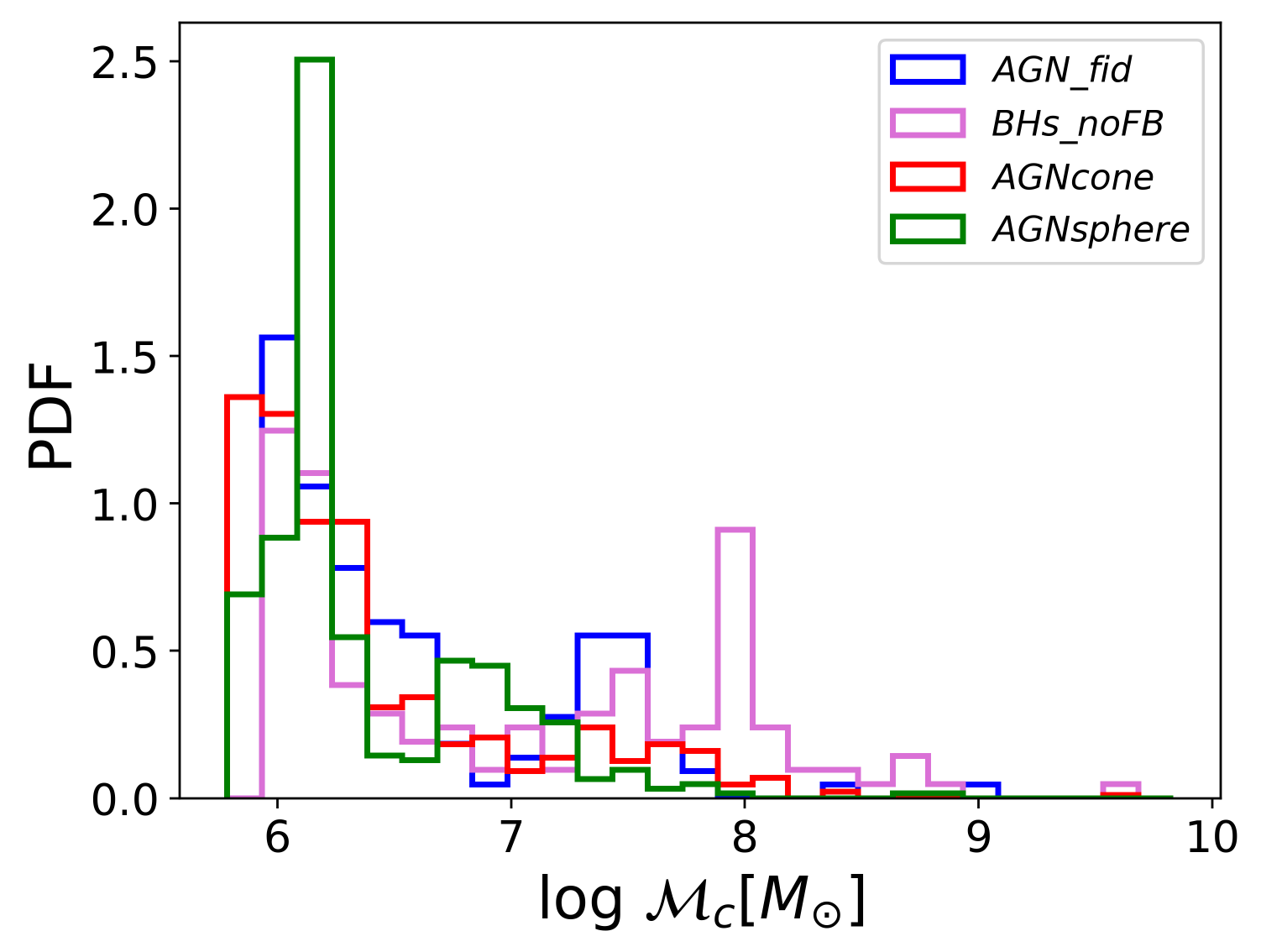}
    \caption{PDF of chirp mass as resulting from the \AGNfiducial{} (blue), \BHsnoFB{} (magenta), \AGNcone{} (red) and \AGNsphere{} (green).}
    
    \label{fig:chirp_mass}
\end{figure}

For the models summarised in Tab. \ref{simuls}, we compute the redshift evolution of the MBH merger rate (per unit redshift, per unit time) for different chirp mass ranges, where the chirp mass is given by ( e.g. \citealt{Cutler_1994,Blanchet_1995}):
\begin{equation}
    %M_c=\frac{-b\pm\sqrt{b^2-4ac}}{2a}.
    { \mathcal M_c}=\frac{(m_1 m_2)^{3/5}}{(m_1+m_2)^{1/5}},
	\label{eq:Mc}
\end{equation}
and $m_1$ and $m_2$ are the masses of the merging black holes\footnote{In these calculations, we do not include those MBHBs that cannot be associated with any galaxy in the simulations. We discuss these spurious events in Sec. \ref{gal-id}}. We show in Fig. \ref{fig:chirp_mass} the normalized probability distribution function (PDF) of the chirp masses resulting from different simulations. Although no evident differences among different models can be seen from this plot, we note that the \citetalias{Barai2018} simulations predict a larger number of small chirp masses ($\mathcal M_c<10^6\msun$) with respect to to \citetalias{Valentini:2021}, and only the \BHsnoFB{} and \AGNcone{} simulations predict MBHBs with $\mathcal M_c>10^9\msun$.

\begin{figure*}
	
	\includegraphics[width=2.0\columnwidth]{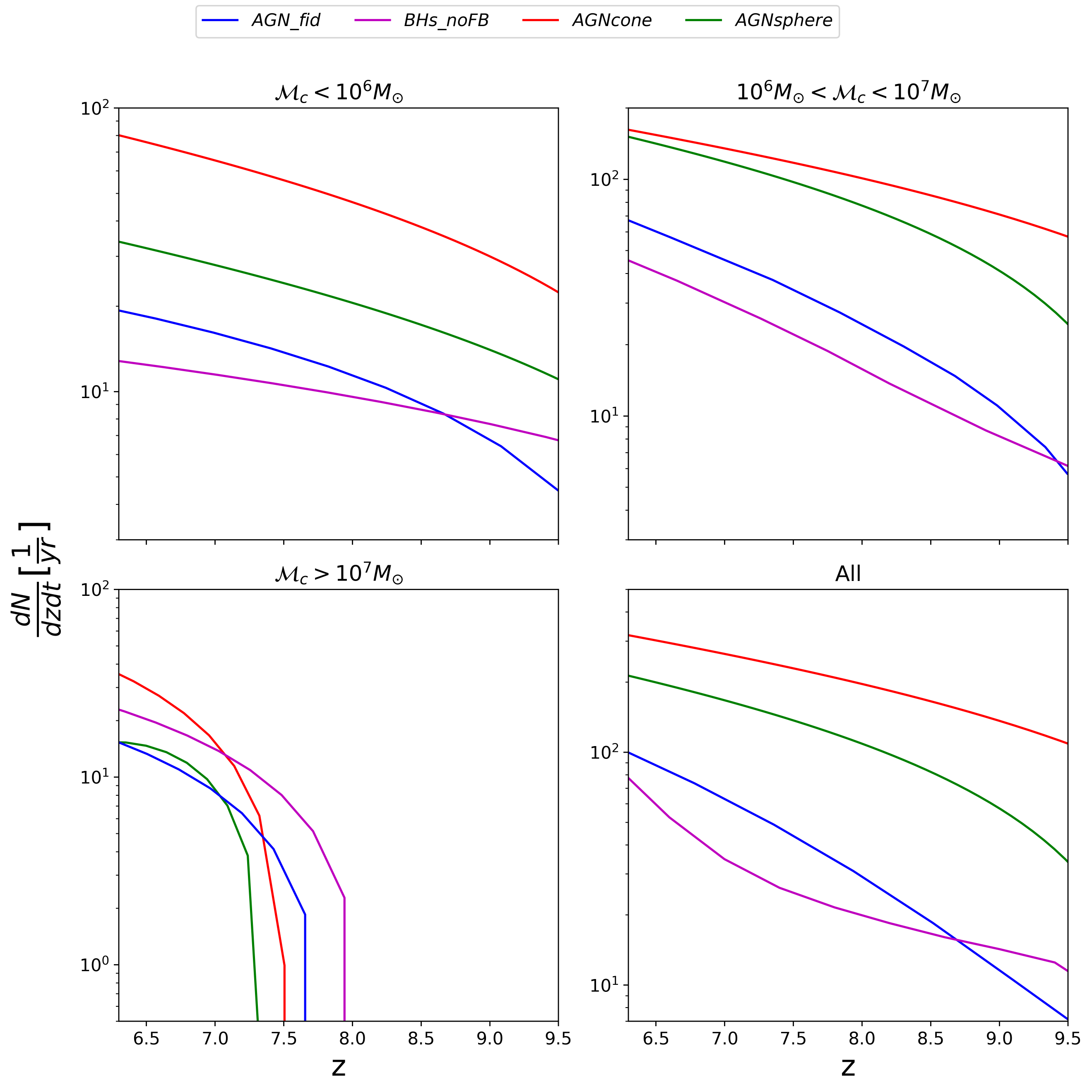}
	\caption{Number of mergers per unit redshift per unit time for \AGNfiducial{} (blue), \BHsnoFB{} (purple), \AGNcone{} feedback (red), and \AGNsphere{} feedback (green). The upper left (right) panel shows the merger rates for MBHB systems with chirp mass $<10^6 \msun$ ($10^6<\mathcal M_c<10^7 \msun$); the lower left panel shows the merger rates for chirp mass >$ 10^7 \msun$, while the lower right panel refers to the the cumulative merger rates for all chirp mass ranges.}
    \label{fig:dN}
\end{figure*}

We then calculate the merger rate from the number density of mergers, dN/dz per comoving volume, dV as (e.g. \citealt{Haehnelt_1994,Ciardi2000}):
\begin{equation}
    %M_c=\frac{-b\pm\sqrt{b^2-4ac}}{2a}.
    \frac{dN}{dz dt_{\rm obs}}=\frac{4 \pi c d_{L}^2 }{(1+z)^2} \frac{dN}{dz dV},
	\label{eq:dN}
\end{equation}
where $d_L$ is the luminosity distance of the event. % and $c$ is the speed of light in vacuum 
The results are shown in Fig.~\ref{fig:dN}.

The number of MBH mergers predicted by our zoom-in cosmological simulations increases with decreasing redshift, as a consequence of the hierarchical structure formation process. Fig.~\ref{fig:dN} shows that the two sets of simulations have different merging histories that depend both on the numerical resolution adopted and on the feedback implemented. 

In particular, Fig.~\ref{fig:dN} shows that the number of mergers predicted by the \citetalias{Valentini:2021} simulations ($\sim$ 100 at $z= 6.5$ for both \AGNfiducial{} and \BHsnoFB{}) is smaller by a factor of at least 2 than the one by the \citetalias{Barai2018} simulations (>200 for both \AGNcone{} and  \AGNsphere{} case at $z\sim 6.5$ ). This major difference is likely linked to the better numerical resolution of the \citetalias{Valentini:2021} simulations with respect to the \citetalias{Barai2018} ones (see table \ref{simuls}). The condition for seeding ($M_h>10^9 \msun$) is more easily satisfied in \citetalias{Barai2018} with respect to \citetalias{Valentini:2021}, because of the larger mass of its resolution elements. This thus implies a larger number of seeds and consequently a larger number of mergers. Even more importantly, the excess of mergers in \citetalias{Barai2018} is driven by the high dynamical mass which enters in the BH-BH merger algorithm within the code.

For a fixed numerical resolution, it is possible to study the dependence of the number of mergers on the feedback implemented. In the \citetalias{Valentini:2021} simulations, we find that for low chirp masses ($<10^7 \rm M_{\odot}$) the merger rate predicted by the \AGNfiducial{} model is higher than the \BHsnoFB{} case.
We further note that for ${\mathcal M_c}>10^7 \rm M_{\odot}$ this trend is reversed, the merger rate in the \BHsnoFB{} run is higher than in the \AGNfiducial{} case. The lack of AGN feedback allows a more efficient gas accretion which makes the black holes to grow more massive and numerous in the \BHsnoFB{} case in this particular chirp mass range.

\begin{figure}
        \includegraphics[width=\columnwidth]{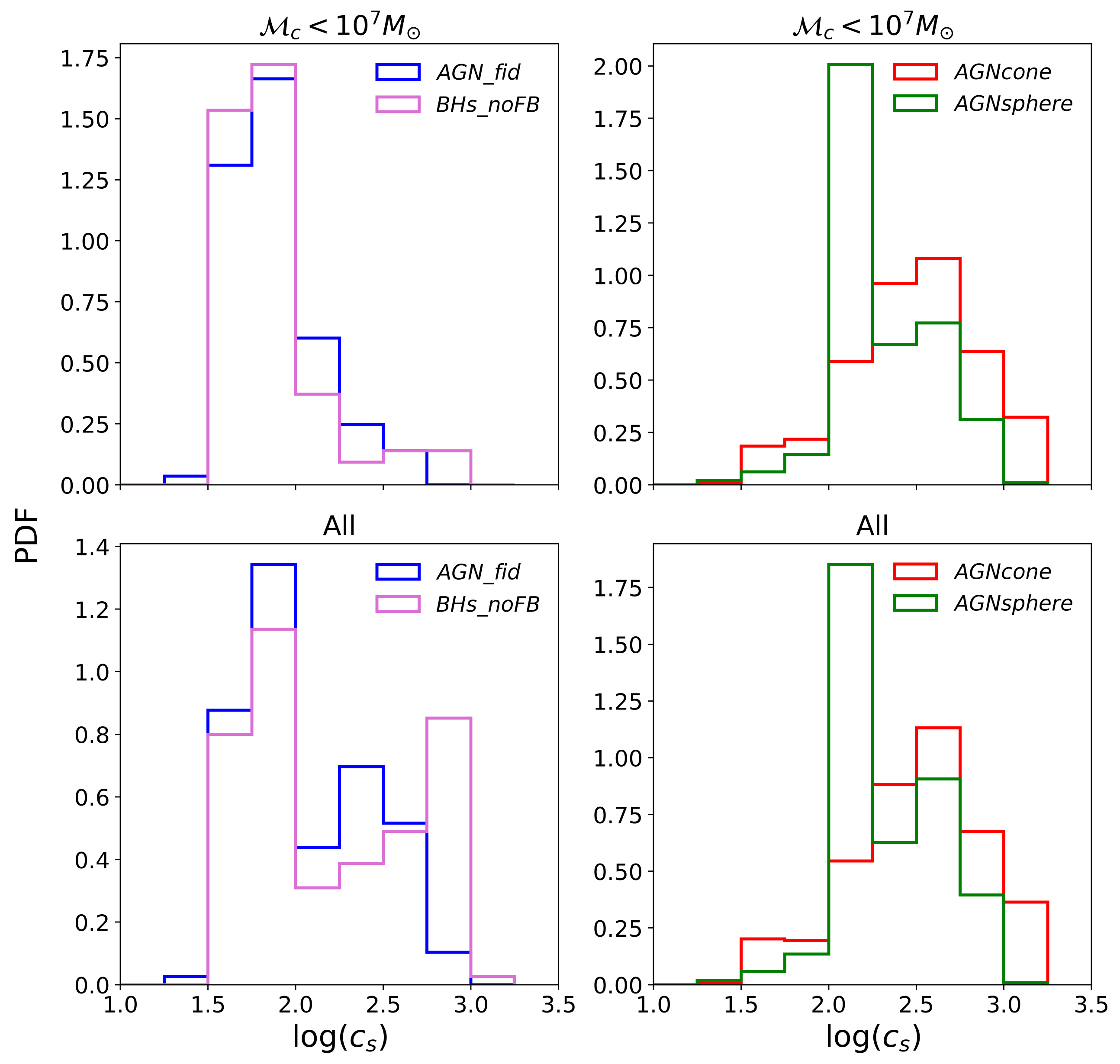}

        \caption{PDF of relative sound speed of merging black holes for \AGNfiducial{} feedback and \BHsnoFB{} from \citetalias{Valentini:2021} (left panel) and \AGNcone{} and \AGNsphere{} feedback from \citetalias{Barai2018} (right panel).The top panels show the distribution of the sound speeds of MBHBs with $\mathcal{M}_c<10^7 \msun$ while the bottom panels show the same for all MBHBs.}
       \label{fig:cs} 
\end{figure}

Furthermore, we study the trend of the local sound speed in the ISM. In the left panels of Fig. \ref{fig:cs} we show the results for the \citetalias{Valentini:2021} simulation runs. We note that the sound speed is larger in the \BHsnoFB{} run because of the following. In the \BHsnoFB{} run, the accretion rate is higher than in \AGNfiducial{} (see Tab. \ref{simuls}); furthermore, on scales over which $c_s$ is computed (i.e. the smoothing length of the black holes), the heating due to accretion (gravitational compression) dominates the heating due to feedback. Thus, the gas temperature (and consequently $c_s$) in the \BHsnoFB{} run is higher than \AGNfiducial{}. We thus conclude that for ${\mathcal M_c} <10^7 \rm M_{\odot}$, the merger rate is driven mainly by the gravity, dynamics and substructure mergers.

For the \citetalias{Barai2018} runs, we observe from Fig.~\ref{fig:dN} that in \AGNcone{} the number of mergers is larger by a factor of $\sim$2 with respect to the \AGNsphere{} case. To investigate this point, in the right panels of Fig. \ref{fig:cs} we show the PDF of the relative sound speed of the merging BHs resulting from these simulations. This figure shows that in the \AGNcone{} case, the PDF is shifted towards larger values. As discussed in Sec. \ref{sec:simulations} two BHs are allowed to merge when their relative velocity is lower than the sound speed of the local ISM. The higher is the sound speed the larger is the probability for two BHs to merge. The sound speed is larger in the \AGNcone{} run because in this case the accretion is by far higher than in \AGNsphere{} (see Tab. \ref{simuls}), resulting in a higher number of mergers.

We further note that \citet{Zana2022} already found that different feedback prescriptions result into different merger rates: in \AGNcone{}, galaxies merge faster and more easily than in \AGNsphere{}, possibly because of the stronger feedback due to the larger black hole accretion rate (see Tab. \ref{simuls}). This determines a more diffuse gas and stellar component around the host galaxies which can boost the effect of dynamical friction when two galaxies approach, thus lowering the dynamical timescale for their merging to occur.

We finally note that different resolutions and feedback prescriptions also affect the epoch at which the furthest merger event is occurring. For example, for ${\mathcal M_c} > 10^7 \msun$, the furthest GW signal occurs in the redshift interval $7.7<z<8$ and $7.3<z<7.5$ in the case of \citetalias{Valentini:2021} and \citetalias{Barai2018} simulations, respectively. 

\subsection{Bias in zoom-in simulations}
\label{sec:halobias}
Our results are based on small box simulations, zoomed-in on a massive, biased \citep[$>3\sigma$, e.g.][]{Barkana_Loeb2001} dark matter halo. As a result, the number density of MBHB mergers within the simulated box overestimates the value in an average region of the Universe.
%This approach systematically overestimates the actual number of MBHB mergers.
In what follows, we estimate the bias of our predictions considering our fiducial run (\AGNfiducial{}).%using \AGNfiducial{} as our typical run. 

Following the halo mass history suggested by \citet{Correa_2015}, we first calculate the accreted mass ($M_{h,0}$) at $z=0$ of a $10^{12}~\msun$ halo at $z=6$:
\begin{equation}
\label{halomass}
M_h(z)=M_{h,0}~(1+z)^{\alpha}e^{\beta z},
\end{equation}
where $\alpha=0.24$ and $\beta=-0.75$, respectively %are parameters arising from the power spectrum of density fluctuations 
\citep{Correa_2015b}. We find $M_{h,0}=10^{13.75}~\msun$.

In practice, when computing the merger rate with Eq. \ref{eq:dN}, we are considering a Universe where all DM halos have $M_h=10^{13.75}~\msun$ and comoving volume $(5.25~\rm cMpc)^3$. The result of this procedure will be clearly biased compared to a proper calculation in which DM halos span a wider mass range ($M_h=10^{10}\sim 10^{16}~\msun$) and have different abundances.
This bias cannot be directly computed from our simulations, but we need to rely on a semi-analytical model (SAM). In SAMs, in fact, a wide range of halos are simulated and the merger rate is computed from their collective output, by weighting each halo mass according to the Press and Schechter halo mass function \citep{PressSchechter_1974}. This latter step is simply obtained by dividing the the merger rate by the effective comoving volume occupied by that halo. It is therefore also possible to use the SAM output to create biased universes, simply by taking halos of a desired mass and weighting them with a desired effective comoving volume. By comparing the rates obtained from this biased universe to the total one, we can infer the bias. This is the procedure we follow here.

We then use the semi-analytical model by \citet[][see also \citealt{Klein_2016}]{Barausse_2012}, and we consider the output of all the trees, weighted on the Press and Schechter halo mass function \citep{PressSchechter_1974}. For these calculations, we consider the "Q3nod" run, since it is based on a model that more closely resembles our prescriptions (heavy seeds and no time delays). Furthermore, for a fair comparison with our results, we only consider $z>6$ merger events occurring in dark matter haloes $>10^9~\msun$ (the threshold mass used in our simulations to seed MBHs) and involving binaries with both BH masses $>10^5~\msun$ (the mass of our seeds). The merger rate obtained in this case is $\sim 3$~yr$^{-1}$.

We next consider two specific merger trees, whose halos at $z=0$ are the closest to the $M_{h,0}$ value computed above, namely $M_{h1,0}=10^{13.7}~\msun$ and $M_{h2,0}=10^{13.8}~\msun$. We weight the merger rates in these two merger trees with the inverse of the comoving volume of our refined simulation, namely (5.25 cMpc)$^3$. The average of the two merger rates obtained in this way can be used as a proxy for the merger rate in a Universe made only of halos of $M_{h,0}=10^{13.75}~\msun$ at $z=0$. The merger rate obtained in this case is $\sim$60~yr$^{-1}$.

We thus estimate that the merger rates computed from our zoom-in simulations is biased by a factor $\sim 20$. We repeat the above calculations for the model "popIII" (which assumes light seeds and delays between MBH and galaxy mergers) and the model "Q3-d" (heavy MBH seeds and delays). In both cases the bias does not change significantly, being $\sim$ 20 to 30.

To summarize, our zoom-in simulations predicts a total number of merger events per year, at $z\sim 6$, that varies between 80 and 300, depending on the resolution and the star formation/AGN feedback prescriptions adopted. By accounting for the halo bias, the number of merger events lower to $\sim$ 3-15~yr$^{-1}$.

\subsection{Comparison with contemporary works}
\label{TC}

In this section, we compare our results at $z=6$ with contemporary works that make similar calculations, both using SAMs and hydrodynamical simulations (HDS). For this comparison, we only consider those models that do not include any time delay in MBHB coalescence, which is consistent with our work\footnote{We further discuss this point in Sec. \ref{sec:delay}.}. %In Sec. \ref{sec:delay}, we make some preliminary calculations about time delay effects}. 
We summarize results from different models in Table~\ref{comp-lit}.

We first compare our results with the predictions by \citet{Sesana_2007} based on the models by \citet{Begelman_2006}. In these works, DCBH formation is efficient when the halos overcome a given threshold of virial temperature ($ T_{\rm vir} \geq 10^4$K): the "high-feedback" (BVRhf) and "low-feedback" (BVRlf) models differ for the efficiency in the distribution of metals produced during the star formation process. In the BVRhf (BVRlf) model, the merger rate is 1.5 (25)~yr$^{-1}$. This difference is due to the following: in BVRhF, as a consequence of the high stellar feedback efficiency that ensures a swift metal enrichment, the DCBH formation stops as early as $z\sim 18$; in the BVRlf, the DCBH formation only stops at $z\sim 15$, since the low stellar feedback efficiency allows halos to remain pristine longer.

\citet{Klein_2016} also investigated the effect of different seeding models and time delays between galactic and MBHB mergers on the merger rate. For this, they adopted the SAM by \citet{Barausse_2012} and varied the seeding mass (light seed vs heavy seed) and hence their halo occupation fraction. They also consider delays in MBHB caused by MBH environment as well as by triple interactions. Considering the model with heavy seeds and instantaneous mergers (model "Q3-nod") at redshift 6, they predict a merger rate of 10 per year. 

We further consider the results by \citet{Hartwig_2018} for instantaneously merging binaries with $10^4<M_{\rm seed}<10^6~\rm M_{\odot}$, as derived by assuming a critical LW flux\footnote{$J_{\rm 21}$ is the LW flux in units of $10^{-21}\rm ergs\: cm^{-2} s^{-1} Hz^{-1} sr^{-1}$.} of $J_c=30 J_{\rm 21}$. In the \citet{Hartwig_2018} calculations, the merger rate the DCBH formation rate peaks at approximately $z\sim 7$ after which the DCBH formation stops, as a result of the metal enrichment and cosmic reionization processes. The resulting merger rate at $z=6$ is 10~yr$^{-1}$.%In the BVRlf model of 

For what concerns \citet{Dayal2019}, they use the SAM of galaxy formation \textit{Delphi} to track the effect of different BH parameters on the BH merger rates. The merger rate at $z = 6$ is found to be $<10$ per year. We only consider the case of instantaneous mergers which doesn't assume any delays between galactic and MBH mergers, referred to as \textit{ins1} model in the \citet{Dayal2019}. We also note that they only report the intrinsic merger rate for all BH mergers (stellar BBH mergers, 'mixed' merger with stellar seed and DCBH as well as DCBH-DCBH mergers). For this reason\footnote{They also show that DCBH mergers, noted as "type 3" mergers, are the rarest in their BH population.}, we consider this estimate as an upper limit to the merger rate when only DCBH are seeded.

For what concerns HDS, \citet{Katz_2019} used \texttt{Illustris} to study different populations of MBHs. They predict the effect on LISA detection rate for different MBH evolutionary scenario such as the effect of delay on the BH particle mergers in the simulations combined with different BH masses. At $z \sim 6$ they predict the intrinsic merger rate  $\sim 0.01$ per year for heavy seeds and instantaneous mergers (model \textit{ND}). 

To summarise, our results are consistent with previous predictions from semi-analytical works (after being corrected for the halo-bias), while they are above the predictions by \citet{Katz_2019}. This inconsistency can be ascribed to the different seeding mechanisms adopted in their work $M_{\rm seed} = 1.42 \times 10^5 \msun$ BHs seeded in $M_{\rm DM} = 7.1 \times 10^{10} \msun$ DM halos): since the DM haloes in which BHs are seeded are more massive (i.e. less numerous) than ours, we expect fewer BHs to be seeded in the simulations, resulting into a lower merger rate. 

Furthermore, in Appendix \ref{PTA}, we compare the gravitational wave background resulting from our simulations with current NanoGrav, EPTA and PPTA observations. We notice that this is only a sanity check to make sure that our predictions are not overshooting current observational constraints. However, it does not represent a genuine comparison since our predictions do not include any contribution from sources at $z<6$, which are instead expected to dominate the background \citep[e.g.][]{Izquierdo_Villalba_2021}.

\begin{table}

\caption{Comparison with contemporary literature for merger rates at $z=6$. The results of our work are reported after being corrected for the halo bias computed in Sec. \ref{sec:halobias}.}
\label{comp-lit}
\centering
\resizebox{0.45\textwidth}{!}{%

\begin{tabular}{|l|l|l|}
\hline
\textbf{Reference} & \textbf{Model} & \textbf{\renewcommand{\arraystretch}{2}\begin{tabular}[c]{@{}l@{}}$\frac{dN}{dz~dt}~ \rm [yr^{-1}]$\end{tabular}} \\ \hline
\citealt{Sesana_2007} & SAM & 1.5-25 \\ \hline
\citealt{Klein_2016} & SAM & 10  \\ \hline
\citealt{Hartwig_2018} & SAM & 5  \\ \hline
\citealt{Dayal2019} & SAM & $<$10 \\ \hline
\citealt{Katz_2019} & HDS & 0.01\\ \hline
This work (bias corrected) & HDS & 3-15 \\ \hline
\end{tabular}%
}

\end{table}
%%%%%%%%%%%%%%%%%%%%%%%%%%%%%%%%%%%%%%%%%%%%%%%%%%%%%%%%%%%%%

\section{Delays in MBHB mergers}\label{sec:delay}
In our simulations, given the limited spatial and temporal resolution, we are not able
%it is impossible 
to properly follow the dynamics of MBHBs up to coalescence. This explains the simplified prescription typically adopted in zoom-in cosmological simulations for BH merging described in Sec. \ref{sec:simulations}. However, the actual timescale over which MBHBs merge depends on several factors, e.g. the mass ratio of the MBHs, their initial separation, and the physical properties of the galaxy hosting the MBHB. In what follows, for our fiducial model, we first describe how we associate a MBHB to its host galaxy, and then we correct in post-processing the coalescing time of MBHB mergers including a time delay due to dynamical friction from the surrounding stars\footnote{In Appendix \ref{stellar hardening}, we describe the stellar hardening physical process that could further delay the coalescence of MBHBs. We find that the resolution of our simulations prevents us to make realistic predictions about this effect.}. This allows us to get a first order estimate of how our merger rates change due to these effects. The final results may anyway vary to some degree from what we report below, if the dynamical friction were actually implemented in the code, instead of applying its effect in post-processing.

\subsection{Galaxy-MBHB association}\label{gal-id}
In this subsection, we describe the method we adopt to associate a MBHB to its host galaxy. Galaxy identification follows a similar approach to what has been done in \citet{Zana2022}. We identify\footnote{To define a halo we require a minimum of 20 bound particles.} dark matter halos through the \textsc{AMIGA} halo finder code \citep{Knollmann2009}. The merger tree for each halo at $z\simeq 6$ is built by tracing back in time the constituent dark matter particles: their ID is matched in the progenitor structures in the previous snaphots.
Baryon particles are assigned to their related galaxy when: ($i$) they are located within $\beta r_{\rm vir}$ of a given halo, where $r_{\rm vir}$ is the virial radius of the halo and $\beta=0.3$; and ($ii$) their velocity is lower than the escape velocity, as evaluated through an analytical integration of the Navarro-Frank-White profile \citep{Navarro1996} to speed up calculations. We restrict our analysis only to those galaxies with $M_{\rm vir} > 10^{9}~\msun$ and $M_{*} > 10^{7}~\msun$.

We associate a host galaxy to each merger event, by using the following procedure:
($i$) we first assign to each merger event that galaxy for which its centre of mass is the closest to the position of the primary BH (BH$_p$); ($ii$) we consider only those mergers which are within $\beta r_{\rm vir}$ of any galaxy.
When we associate the galactic properties (derived from a snapshot) to a merger event, we consider the closest\footnote{We have also considered the case in which galactic properties are linearly interpolated from the two snapshots immediately before and after the merger event, finding no appreciable differences in the main results of our work.} snapshot in time to the event itself (see Sec. \ref{sec:discussion} for further discussion).

In some cases, our algorithm fails to associate a host galaxy to a merger event. This may occur because during the time passing between the redshift of the event and the closest snapshot, the BH$_p$ may have moved out of the $r_{\rm vir}$ of the host galaxy. Another possibility is related to the fact that during the simulation a MBH may be spuriously seeded into a transient matter overdensity (incorrectly identified as a galaxy by the on-the-fly halo finder); such a MBH would then rapidly merge with the MBH of the closest halo (due to the repositioning algorithm) in less than 1 timestep (e.g., \citealt{Blecha_2015, Kelley_2016, Katz_2019}). 

We find, in our fiducial model, over the 145 total events, 72 per cent occur within $\beta r_{\rm vir}$, 22 per cent outside $\beta r_{\rm vir}$ but inside $r_{\rm vir}$, and 6 per cent outside the $r_{\rm vir}$. We remove these spurious events from our calculations.\footnote{ We also calculate the spurious events in \AGNcone{} for comparison of different resolutions and we find that the fraction of spurious events increase with the decrease of resolution of the numerical simulation. Over the 1812 total events in \AGNcone{}, only 10 per cent occur within $\beta r_{\rm vir}$, 20 per cent outside $\beta r_{\rm vir}$ but inside $r_{\rm vir}$, and 70 per cent outside the $r_{\rm vir}$. This further strengthens our selection of \AGNfiducial{} %from the higher resolution simulation suits 
 as our fiducial model.}
\begin{figure*}
    \includegraphics[width=2.1\columnwidth]{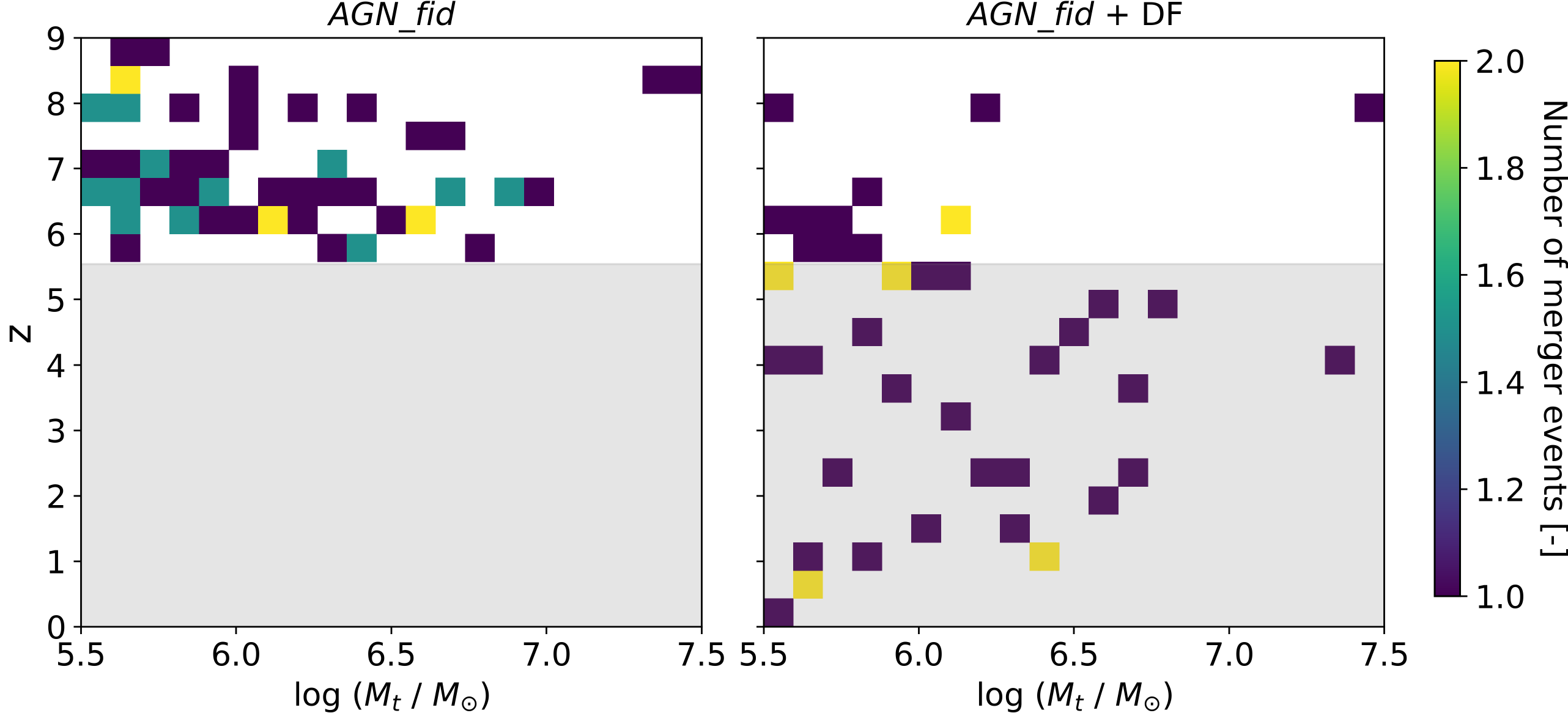}

    \caption{Number of merger events as a function of the total mass of the system $M_{t}$ and the redshift $z$ of the binary, as resulting from our fiducial run. The left panel shows the merger distribution resulting from the simulations while the right panel shows the same distribution after adding the delay time due to dynamical friction. The white region in each plot shows the redshift range accessible to the simulations ($z$>6) and the grey region denotes the redshift range inaccessible to our simulations. }
    \label{fig:delay}
\end{figure*} 
\subsection{Time delay due to dynamical friction}
The interaction of the MBHBs with stars in their surroundings results in the MBHs to lose energy, to slow down, and to spiral inwards gradually. \citep{Chandrasekhar1943, Ostriker_1999}. This process effectively increases the timescale of the MBHB merger with respect to the adopted simulations, potentially delaying it by millions or even billions of years. The amount of dynamical friction experienced by MBHBs depends on the density and distribution of the surrounding stars, as well as the mass and velocity of the binary. In general, the effect of dynamical friction is strongest in regions of high density, such as the centers of galaxies, where the density of dark matter and stars is the highest. 
We make a simple calculation following the prescription of \citet{Krolik_2019}, but see also \citet{Volonteri2020} for more details. 

    The frictional timescale for a massive object in an isothermal sphere can be written as \citep{Binney2008}:
\begin{equation}
t_{\rm df}=0.67\, {\rm Gyr} \left(\frac{a}{4\, {\rm kpc}}\right)^2\left(\frac{\sigma}{100 \, {\rm km\, s^{-1}}}\right)\left(\frac{M_{\rm BH_s}}{10^8 \,M_{\odot}}\right)^{-1}\frac{1}{\Lambda},
\label{eq:tdf}
\end{equation}
where $a$ is the distance of the MBH from the galaxy centre\footnote{We calculate $a$ at the snapshot closest in time to the numerical merger.}, $\sigma$ is the central stellar velocity dispersion:
\begin{equation}
\sigma= (0.25GM_{*}/R_{\rm eff})^{1/2},
\label{eq:tsig}
\end{equation}
$M_{*}$ is the total stellar mass of the galaxy hosting the MBHs, computed as described in Sec. \ref{gal-id},
\begin{equation}
\Lambda=\ln(1+M_{*}/M_{\rm BH_s}), 
\label{eq:tlam}
\end{equation}
$R_{\rm eff}$ = 0.1 $r_{\rm vir}$, and $M_{\rm BH_s}$ denotes the mass of the secondary (less massive) MBH. 

The total time taken by the MBHBs to merge including the dynamical friction correction is then given by:
\begin{equation}
t_{\rm tot,df}= t_{\rm in} + t_{\rm df},
\label{eq:tdftot}
\end{equation}
where $t_{\rm in}$ is the time at which the merger occurs in the simulation.\\

\begin{figure}
    \includegraphics[width=0.9\columnwidth]{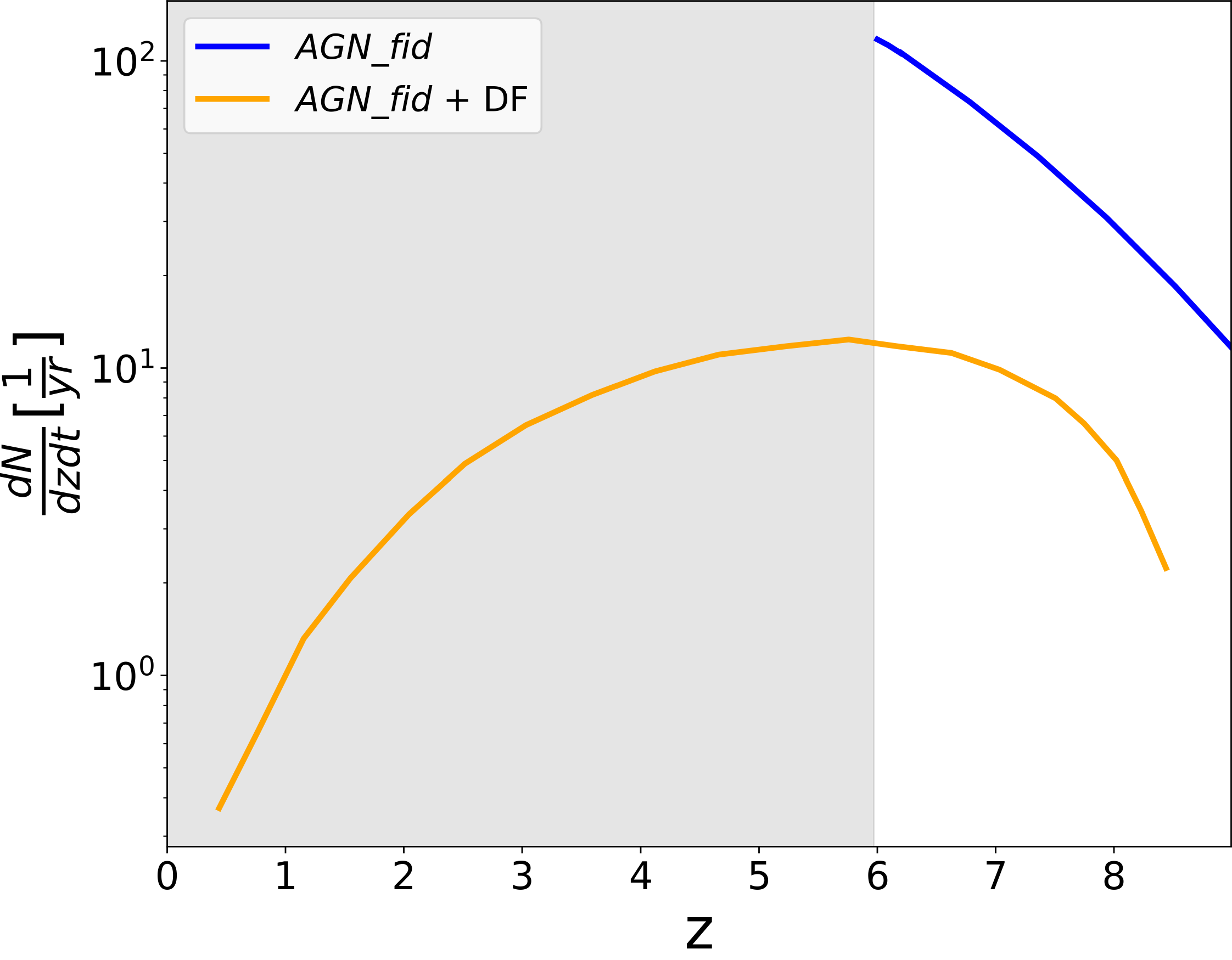}
    \caption{Merger rate per unit redshift $z$ of MBHBs, considering delay times in post-processing. In blue, mergers from \AGNfiducial{} are shown and the same events which can be associated with a host galaxy are shown in orange after adding the delay time due to dynamical friction. The gray shaded region depicts the redshift range which is inaccessible in our simulations. }
    \label{fig:mrdelay}
\end{figure}

The results of our calculations are reported in Fig. \ref{fig:delay} where we show the number of merger events (for which we can associate a host galaxy) across different mass and redshift ranges without including any post-processing delay (left panel), and including dynamical friction (right panel). The grey region in each panel denotes the redshift range outside the reach of our simulations ($z$ < 6). Fig. \ref{fig:mrdelay} helps to better visualise the difference in the merger rate predictions if we assume instantaneous merger (blue line), or we include delay due to dynamical friction (yellow line) in post-processing. By adding the delay due to dynamical friction, we find that 17 per cent of the MBHBs of our fiducial calculations are not merging within the Hubble time, 21 per cent of the MBHBs merge at $z > 6$ and the rest will be delayed to a redshift range $z < 6$.

\section{ GWs from high redshift MBHBs}\label{sec:GWhighz}
In this section, we estimate the detectability of the merger events predicted by the \citetalias{Valentini:2021} \AGNfiducial{} simulations\footnote{In Appendix\ref{GW_all}, we compare the GW properties reported in this section for the \AGNfiducial{} with the other simulation runs presented in Sec. \ref{sec:simulations}.}, by computing the signal-to-noise ratio (SNR) and the angular resolution $\Omega$ of their GW signals. We also show how these properties vary if time delay due to dynamical friction is considered. We assume that a GW signal is detectable if SNR$ > $5. Hereafter, for the sake of brevity, we refer to these "LISA detectable events" as LDEs. Furthermore, we also depict the fraction of LDEs in the "mass ratio"-"total mass" plane with and without considering delay effects. Finally, we discuss the different scales of interest in time, frequency and spatial ranges of the LDEs.

\subsection{Signal-to-noise ratio and angular resolution}\label{snrtext} 

The SNR accumulated over the observational time $\tau$ is computed following the \citet{Flanagan_1998} formalism:
\begin{equation}
\left (\frac{S}{N}\right)_{\Delta f}^2= \int_{f}^{f+\Delta f} d\ln f' \, \left[
\frac{h_c(f'_r)}{h_{\rm rms}(f')} \right]^2,
\label{eqSN}
\end{equation}
where $f_r$ is the GW rest-frame frequency, $f=f_r/(1+z)$ is the observed frequency, $\Delta f$ is the frequency
shift in the duration of $\tau$, $h_c$ is the {\it characteristic strain}, and $h_{\rm rms}$ is the effective\footnote{The total LISA $h_{\rm rms}$ noise is the sum in quadrature of the instrumental rms noise and the confusion noise from  unresolved galactic \citep{Nelemans}, extragalactic \citep{Farmer_2003}, and white dwarf-white dwarf binaries. The number of these sources is expected to decrease as the LISA mission progresses and a larger number of foreground sources are detected and removed.} rms noise of the instrument. 

\begin{figure}
   \centering
   \includegraphics[width=\columnwidth]{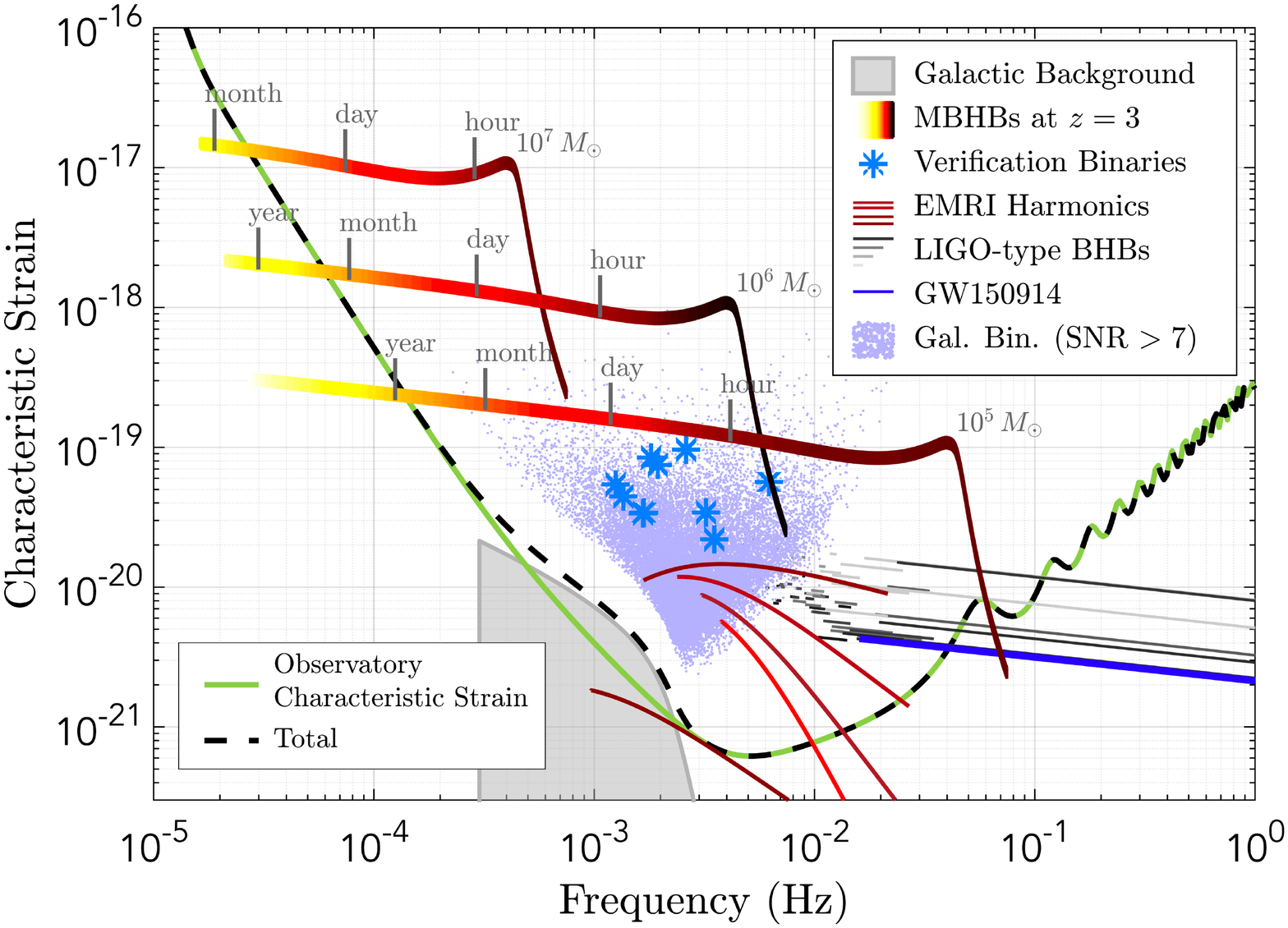}
   \caption[Caption for LOF]{The yellow-red lines shows the varying characteristic strains through the inspiral, merger and ringdown phases for MBHB systems of total masses $10^5,10^6,10^7 \msun$ from bottom to top.  Expected sensitivity (green) with various possible sources in units of dimensionless characteristic strain amplitude for a three arm configuration of LISA. Plot taken from the LISA L3 mission proposal.\protect\footnotemark}
   \label{fig:lisa}
\end{figure}

We start defining the strain amplitude (sky and polarization averaged) of GWs emitted by two black holes of chirp mass ${\mathcal M_c}$ that are merging at redshift $z$, following \citet{Hawking1989}: 
\begin{equation}
    h\,=\,\frac{8\pi^{2/3}}{10^{1/2}}\,\frac{G^{5/3}{\mathcal M_c}^{5/3}}{c^4r(z)}\,f_r^{2/3},
	\label{eq:h}
\end{equation}
where $r(z)$ is the luminosity distance of the merging events. MBHBs spend a mass-dependent amount of time in each frequency band, as shown in Fig.~\ref{fig:lisa}.
\footnotetext{\url{https://www.elisascience.org/files/publications/LISA_L3_20170120.pdf}}

It is thus common to compute the characteristic strain amplitude that also depends on the number of cycles spent in the LISA bandwidth by the binaries:

\begin{equation}
    h_c=h\sqrt{n} \simeq \frac{1}{3^{1/2}\pi^{2/3}}\,\frac{G^{5/6}
{\mathcal M_c}^{5/6}}{c^{3/2}d_L}\,f_r^{-1/6},
	\label{eq:hc}
\end{equation}
where $n$ is the number of cycles spent in a frequency interval  $\Delta f $:
\begin{equation}
n \simeq f_r^2/\dot{f_r}=\frac{5}{96\pi^{8/3}}\,\frac{c^5}{G^{5/3}{\mathcal M_c}^{5/3}}\,
f_r^{-5/3},
\label{eqenne}
\end{equation}
and the rest-frame frequency shift rate is expressed as:
\begin{equation}
\dot f_r=\frac{df_r}{dt_r}= \frac{96\pi^{8/3}G^{5/3}}{5c^5}{\mathcal M_c}^{5/3}f_r^{11/3},
\label{eqdotf}
\end{equation}
assuming that the backreaction from GW emission dominates the orbital decay of a binary.
\begin{figure*}

	\includegraphics[width=1.6\columnwidth]{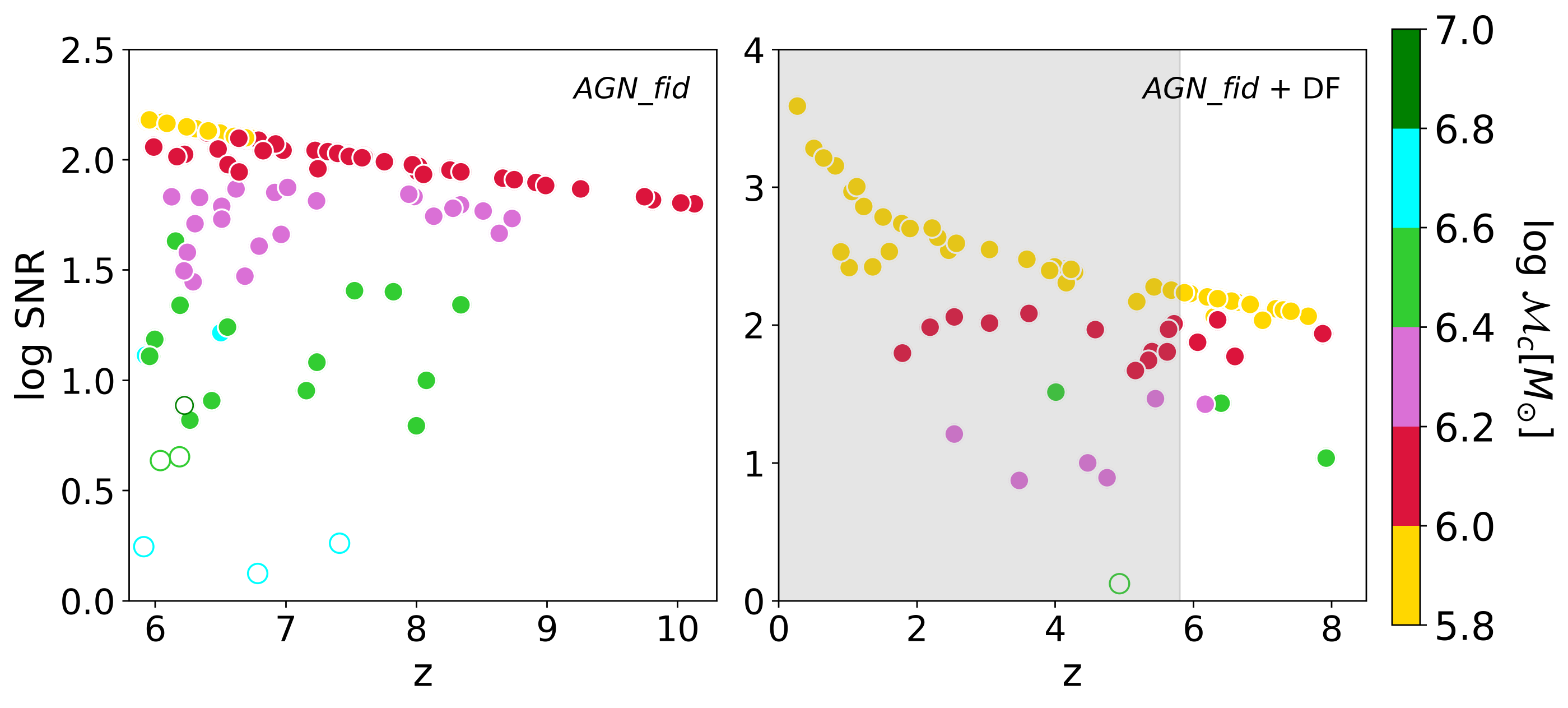}
	\includegraphics[width=1.6\columnwidth]{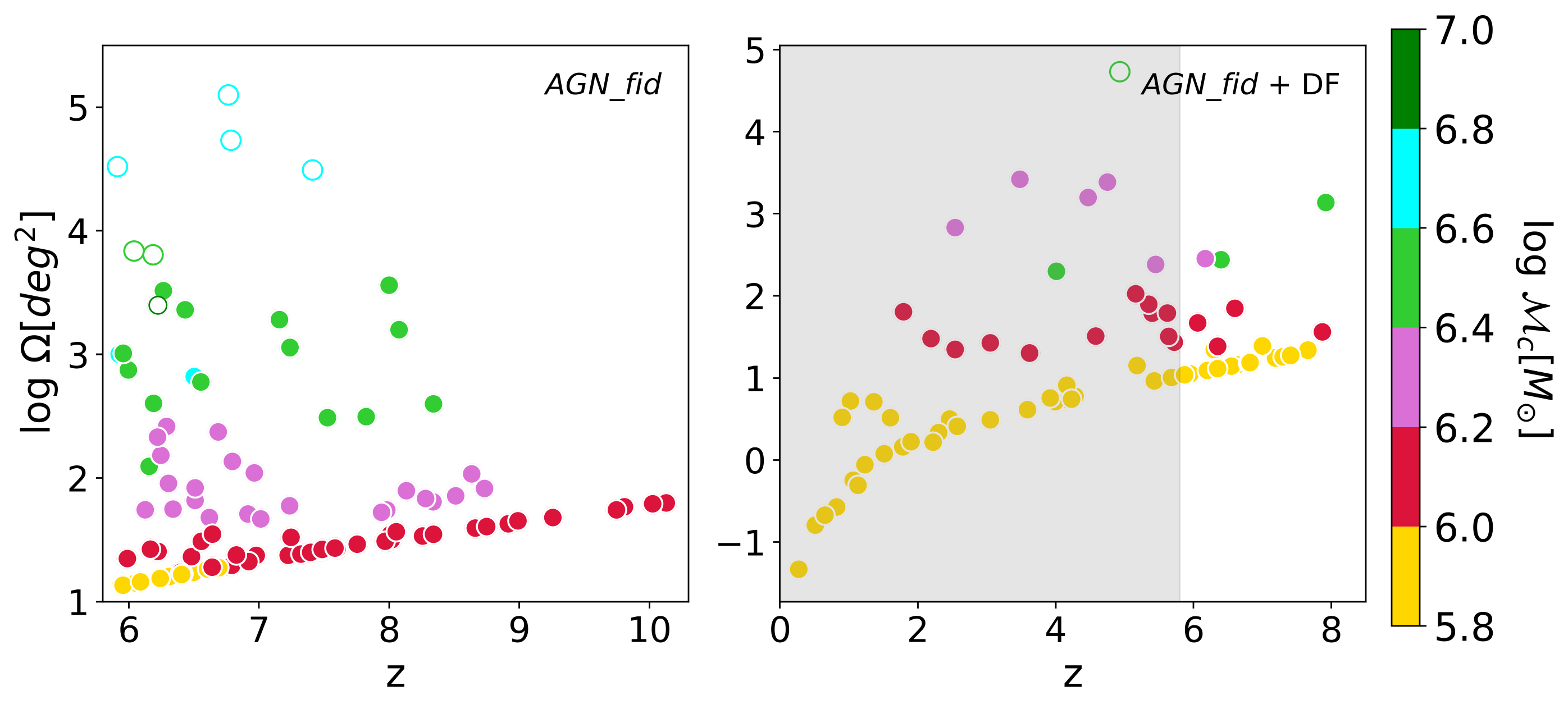}
    
    \caption{ Signal-to-noise ratio (upper panel) and angular resolution (lower panel) of MBHB mergers resulting from our 
    simulations, with (without) delay times considered in right (left) panel color-coded according to their chirp mass. The shaded gray area represents the redshift range inaccessible to our simulations. Filled and empty circles represents detectable and undetectable events, respectively. The detectability threshold has been set to SNR$_{\rm thres}$ = 5.}
    \label{fig:snr}
\end{figure*}

The LISA rms noise $h_{\rm rms}$ is instead given by:
\begin{equation}
   h_{\rm rms}(f)\equiv \sqrt{\Delta f S_n(f)}, 
\end{equation}
where $S_n$ is the LISA power spectral density (PSD): 
\begin{equation}\label{PSD}
\begin{split}
    S_{n}(f)&=\frac{20}{3}\frac{4S_{\rm n,acc}(f)+S_{\rm n,sn}(f)+S_{\rm n,omn}(f)}{L^{2}}\\
    &\times \left[1+\left(\frac{f}{\frac{0.41c}{2L}}\right)^{2} \right]\,,
    \end{split}
\end{equation}
$L$ corresponds to the detector arm length, and $S_{\rm n,acc}$, $S_{\rm n,sn}$ and $S_{\rm n,omn}$ are the noise components due to low-frequency acceleration, shot noise and other measurement noise, respectively  (\citealt{Smith_2019,Klein_2016}), parametrised as follows\footnote{These values hold for the current LISA design, that presents three spacecrafts connected by 6 links.
Given the large uncertainties
on the very-low frequency LISA sensitivity, we adopt a pessimistic cut
at $10^{-4}$ Hz \citep{Sesana_2004}.}:
\begin{equation}
    \begin{split}
        &S_{\rm n,acc} =\frac{9\times10^{-30}}{(2\pi f)^{4}}\left(1+\frac{10^{-4}}{f}\right)\,\,[\rm{m^{2}Hz^{-1}}],\quad \\
        &S_{\rm n,sn} =2.22\times 10^{-23}\,\,[\rm{m^{2}Hz^{-1}}],\quad \\
        &S_{\rm n,omn} =2.65\times 10^{-23}\,\,[\rm{m^{2}Hz^{-1}}].
    \end{split}
\end{equation}

In an ideal experiment, to maximize the SNR, one should integrate eq. \ref{eqSN} over the entire duration of the GW event. Most of the lifetime of a GW emitted by a MBHB is encompassed within the time interval between when the distance between the two MBHs becomes close to the {\it hardening} radius $r_h$ (the inspiral phase begins) and when it reaches the {\it innermost-stable circular  orbit} radius $r_{\rm isco}$ (the merging phase begins). In this case, the integration limits should range between a minimum frequency $f_{\rm min}=f_h$ at the $r_h$ and the frequency $f_{\rm max}=f_{\rm isco}$ at the $r_{\rm isco}$. However, in a real experiment, a GW event can be detected by LISA only if its frequency is included in the range [10$^{-4}$--1.0]~Hz and the SNR overcomes a certain threshold (here taken as SNR$_{\rm thresh}$ = 5). We thus consider as $f_{min}$ the frequency at which SNR$ > $SNR$_{\rm thresh}$. We hence calculate the SNR for each merger event in \AGNfiducial{}, and we find the results shown in the upper panels of Fig. \ref{fig:snr}. 

The upper left panel of fig. \ref{fig:snr} shows the trend of the SNR with redshift, without assuming any time delay. For a given chirp mass, the further the source is located, the lower is the SNR of the GW event. Furthermore, although $h_c$ increases with the mass of the MBHB system, for a fixed redshift, the SNR is higher for sources with lower chirp mass. This trend occurs because low-mass binaries merge slower and enter in the LISA band sooner: hence, they stay in the LISA band for longer time and accumulate more SNR over their inspiraling lifetime. The maximum SNR that is resulting in this case is $\sim 100$.

The upper right panel of fig. \ref{fig:snr} shows instead the same trend of the SNR with redshift, when time delay due to dynamical friction is taken into account. We find that in this case events delayed at epochs $z < 6$ are characterized by SNRs that can be as high as $10^3-10^4$, while at $z>6$ the highest SNR limit remains the same as \AGNfiducial{} without any delays (SNR $\sim$ 150). 

\begin{table*}
\begin{center}
\caption{Detectability of MBHBs with LISA for the simulation runs adopted in this work (first column). The second column reports the total number of mergers in each run ($N_{\rm total}$) while the third shows the LDEs fraction ($f_{\rm det}$) in each run. The fourth (fifth, sixth, seventh) column the chirp mass (observational time, observed frequency and observed initial distace of separation) ranges for the LDEs. The eighth (ninth)column shows the fraction $f_{\rm out}$ ($f_{\rm und}$) of MBHBs whose frequency is outside the LISA band (inside the LISA band but do not reach the required SNR threshold). In the last column the minimum and maximum SNR accumulated by MBHBs which merge within the LISA band but fail to reach the SNR threshold for detectability (SNR$_{\rm und}$) are also shown.}
	\label{lisatab}
\resizebox{\textwidth}{!}{%
\begin{tabular}{|c|c|c|c|c|c|c||c|c|c|}
\hline
\textbf{Run} & \textbf{$N_{\rm total}$} & \textbf{$ f_{\rm det}$} & \textbf{${\mathcal M_c}[10^{6}\msun] $} & \textbf{$ t_{\rm obs}\rm[days]$} & \textbf{$ \nu_{\rm obs}\rm[mHz]$} & \textbf{$ R_{\rm obs}\rm[10^{-6}pc]$} & \textbf{$ f_{\rm out}\rm[mHz]$} & \textbf{$ f_{\rm und}\rm[mHz]$} & \textbf{$SNR_{\rm und}$} \\ \hline
\textbf{\AGNfiducial{}} & 145 & 0.69 & 0.9-7.0 & 0.5-30 & 0.10-0.38 & 1.9-9.3 & 0.31 & 0.04 & 0.8-4.0 \\ \hline
\textbf{\AGNfiducial{}+DF} & 116 & 0.66 & 0.2-3.2 & 0.0-18 & 0.10-0.9 & 0.1-9.6 & 0.31 & 0.01 & 1.33 \\ \hline
\textbf{\BHsnoFB{}} & 140 & 0.46 & 0.9-6.0 & 0.7-39 & 0.09-0.38 & 1.9-9.5 & 0.50 & 0.03 & 0.7-3.9 \\ \hline
\textbf{\AGNcone{}} & 583 & 0.76 & 0.6-10 & 1.1-75 & 0.09-0.42 & 1.8-10.5 & 0.22 & 0.02 & 0.4-4.7 \\ \hline
\textbf{\AGNsphere{}} & 415 & 0.78 & 0.6-7.5 & 0.6-54 & 0.09-0.40 & 1.6-8.4 & 0.22 & 0.005 & 1.6-3.7 \\ \hline
\end{tabular}
}
\end{center}
\end{table*}

Finally, for the prospect of follow-up observations with electromagnetic telescopes, we calculate the LISA angular resolutions of LDEs to quantify how large is the region in the sky that must be covered by a telescope to detect the electromagnetic signals from merging MBHBs. We adopt the results found by \citet{mcgee2020linking}, which are derived from a range of population models: the median angular resolution ${{\Omega}}$ can be associated to the median SNR at which a merger is observed by the following relation: 
\begin{equation}
  {\Omega}\approx 0.5\,\left(\frac{\rm SNR}{10^3}\right)^{-7/4}  {\rm deg}^2\,.
  \label{eq:angres}
\end{equation}

The lower panel of Fig. \ref{fig:snr} tracks the redshift evolution of the angular resolution for LDEs. Following the SNR trend, the optimal angular resolution is found for lower redshift (higher $h_c$) and smaller chirp mass (higher SNR) systems. This figure clearly shows that, although GW events from MBHB coalescence can be detected at high-$z$, their sky localization is poor \citep[10 deg$^2$ in the most optimistic case; see also][]{McWilliams_2011}, making follow-up observations in different EM bands challenging. 
\subsection{Mass ratio}
\label{reobs}

\begin{figure}
	
        \includegraphics[width=\columnwidth]{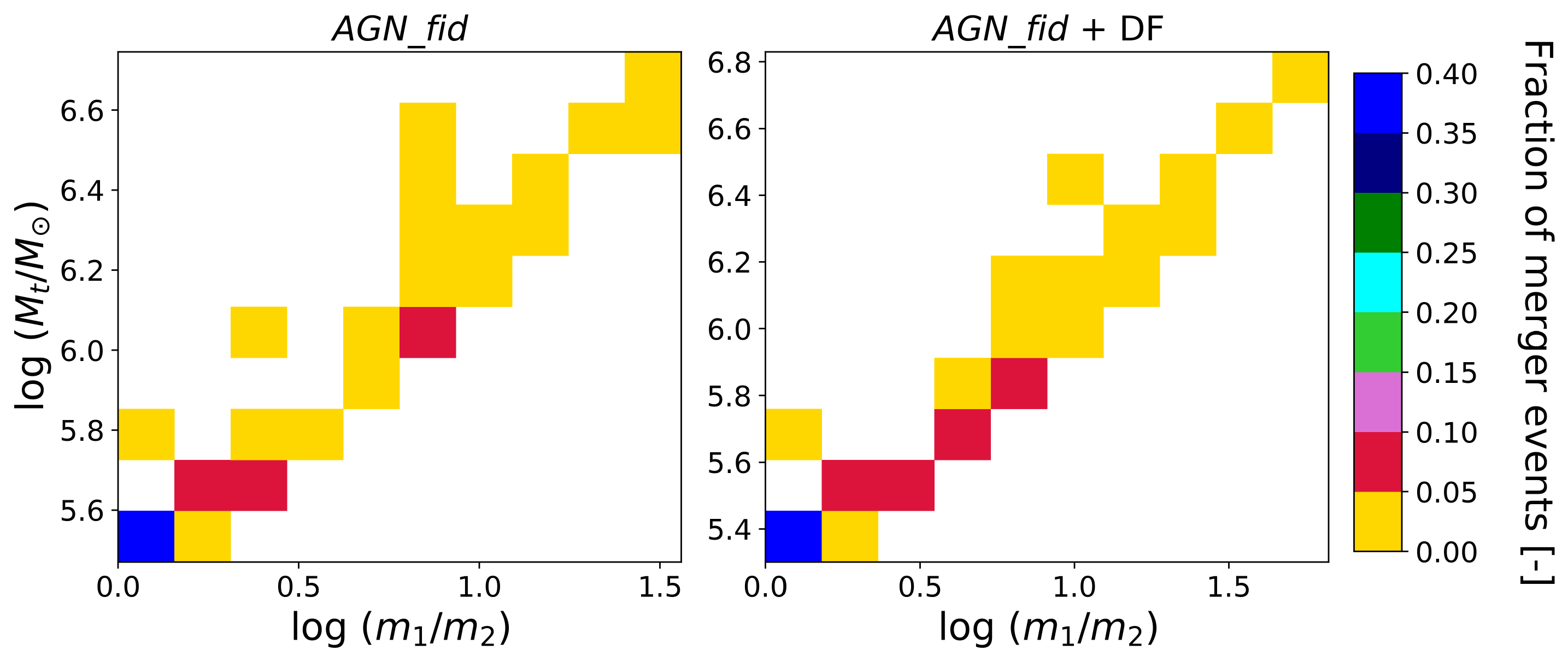}
    \caption{Fraction of detectable merger events as a function of the total mass of the system $M_t=m_1+m_2$ and the mass ratio $m_1/m_2$ of the binary.}
    \label{fig:massratio}
\end{figure}

Fig. \ref{fig:massratio} shows how the number of LDEs is distributed in terms of the "mass ratio" (defined as $m_1/m_2$, and shown in the x axis) and the "total mass" ($M_t=m_1+m_2$, and shown in the y axis), for \AGNfiducial{} with (right) and without (left) considering additional delay in the mergers due to dynamical friction. 

The highest fraction of LDEs occurs for equi-mass binaries, in the low mass range ($M_t\lesssim 3 \times 10^{6} \msun$), irrespective of consideration of delay time and is about 35 per cent to 40 per cent of the total LDE population in each run. In other words, "just-seeded" black holes in binaries have a higher probability to coalesce as compared to black holes with higher mass, assembled by accretion and/or merging. 

Furthermore, we note that for a fixed $M_t$, the larger is the mass ratio the higher is the fraction of mergers. This is simply due to the fact that the number of black holes decreases with increasing masses, thus for a fixed $M_t$ most of the mergers occurs for the BHs with lowest mass.  

\subsection{Time, frequency, and spatial scales of interest}
Two MBHs\footnote{We only consider binary coalescences in our analysis, neglecting systems composed by three (or more) MBHs.}  at a distance $R$ require a certain time to merge through the emission of GWs. Such a time scale is called {\it coalescing} time ($t_{\rm coal}$) and defined as follows: 
\begin{equation}
    t_{\rm coal}=\frac {5}{256 }\frac{c^5 R^4}{G^3 M_{t}^2 \mu},
	\label{eq:t}
\end{equation}
where $R$ is the initial separation of the two merging MBHs and $\mu=M_1M_2/M_{\rm t}$ is the symmetric mass ratio. However, as already mentioned in Sec. \ref{snrtext}, we can detect the GW signal only after SNR$>$SNR$_{\rm thresh}$. We thus compute the {\it observational} time as the time interval between the moment when SNR$>$SNR$_{\rm thresh}$ and when the two MBHs start merging:
\begin{equation}
   t_{\rm obs}= t_{\rm coal}\vert_{\rm SNR > SNR_{\rm thresh}},
   	\label{eq:tobs}
\end{equation}
which provides the period of time during which the GW event is actually observable. This timescale provides the interval of time required to eventually trigger electromagnetic telescopes for follow-up observations to find the EM counterpart of the MBHB merging \citep[see for example][]{Loeb_2016}, or to simply probe the host galaxies of the system.

To the observational time $t_{\rm obs}$ of a merger event we can associate the frequency $\nu_{\rm obs}$ and the separation of the MBHs such that SNR$>$SNR$_{thresh}$, which are related by the following expression:
\begin{equation}
   \nu_{obs}=\pi^{-1}\left(\frac{G M_t}{R_{obs}^3}\right)^{1/2}.
\end{equation}
We summarise the resulting scales of interest in Table \ref{lisatab}.

\section{Caveats and discussion}
\label{sec:discussion}
The results presented in this work are affected by several limitations:
\begin{itemize}
    \item Spatial resolution: the dynamical range required to study galaxy formation and BH co-evolution through numerical simulations is extremely large, ranging from hundreds of Mpc scales (to search for massive DM halos in N-body simulations) to sub-kpc scales (required to properly model gas accretion onto BHs and the subsequent feedback processes). This implies that the spatial resolution that is possible to achieve within a reasonable amount of computational time is typical limited to tens of pc scale\footnote{The largest spatial resolution achieved so far by zoom-in cosmological simulations is $\sim$15 pc of a z = 7 quasar \citep{Lupi2019}.}, namely more than 7 orders of magnitude larger than the typical scales of interest for GW studies (see Table \ref{lisatab}). 
    \item Temporal resolution: a discussion similar to the spatial resolution can be done for what concerns the temporal resolution. Black hole properties in our simulations are evolved with time-steps of $\sim 0.01-1$ 
    Myr, which are several orders of magnitude larger than the typical scales of interest for GW studies (see Table \ref{lisatab}). The situation gets worse if we consider the time interval between two snapshots ($\sim$ tens Myr). This implies that we are not associating an LDE to the galaxy properties at the time of the coalescence (see Sec. \ref{gal-id}). 
    \item Seeding prescription: our simulations are based on a seeding prescription that mimicks the DCBH, given the mass of the seeds that we choose. However, the formation of a DCBH is governed by a complex network of physical processes (i.e. H$_2$ formation, metal enrichment, radiative transfer) that is impossible to take into account self-consistently in cosmological zoom-in hydrodynamical simulations. Our seeding prescription (namely a BH of mass $10^5 \msun$ in each $10^9 \msun$ DM halo) is thus certainly overestimating the number of MBHs in \citetalias{Barai2018} simulations \citep[see e.g.][]{Vito2022} and consequently the expected number of merger events reported here.

\end{itemize}

For the reasons reported above, the merger rate computed in this work represents a solid {\it upper limit} on the number of merger events that LISA will be able to detect. 

\section{Summary and conclusions}
\label{sec:conclusions}
In this work, we adopted a suite of cosmological zoom-in hydrodynamical simulations of galaxy formation and BH co-evolution, developed with the GADGET-3 code, and characterised by different numerical resolutions and star formation/AGN feedback prescriptions. Our simulations are based on a seeding prescription such that $M_{\rm seed} \sim 10^5 \msun$ BHs are planted in $M_{\rm DM} \sim 10^{9} \msun$ DM halos; furthermore, if two BHs are at a distance smaller than the smoothing length of the simulations and their relative velocity is smaller than the local sound speed, they are assumed to merge instantaneously. We use these simulations to investigate the coalescence of massive black hole ($M_{\rm BH}\gtrsim 10^{6}~\rm M_{\odot}$) binaries at $6<z<10$ and to compute their GW properties. We summarise below the findings of our work and draw the main conclusions arising from these results.

\begin{itemize}
\item  {\bf Merger rate :} 
We calculated the merger rates of MBHBs with different AGN feedback scenarios and numerical resolution: thermal (\AGNfiducial{} and \BHsnoFB{} by \citetalias{Valentini:2021}, $m_{\rm DM}=1.5\times 10^6~\msun$) and kinetic (\AGNcone{} and \AGNsphere{} by \citetalias{Barai2018}, $m_{\rm DM}=7.5\times 10^6~\msun$). We found that the merger rate strongly depends on the numerical resolution adopted, ranging within 80 to 300 %20 and 200 
events per year in the \citetalias{Valentini:2021} and \citetalias{Barai2018}, respectively at redshift 6. Furthermore, the merger rates of MBHBs at fixed resolution depend on the feedback recipe implemented: the \AGNfiducial{} model predicts a merger rate that is a factor $\times$ 2 higher than \BHsnoFB{} at $z\sim 6$ for chirp masses in the range $10^6<\mathcal M_c<10^7~\msun$; analogously, the \AGNcone{} model predicts a merger rate that is a factor $\times$ 3 higher than \AGNsphere{} at $z\sim 6$ both for low ($\mathcal M_c<10^6~\msun$) and high ($\mathcal M_c>10^7~\msun$) chirp masses. Different feedback prescriptions and numerical resolutions also affect the epoch of the furthest GW signal detectable with LISA: in the \citetalias{Valentini:2021} simulations, for ${\mathcal M_c} > 10^7 \msun$, the furthest GW signal occurs at $z\sim 7.7-8.0$ while in the \citetalias{Barai2018} at $z\sim 7.3-7.5$. We discussed in details all the physical and numerical explanations for these trends and we underlined the several motivations that make our predictions stringent upper limits to the actual merger rates that are expected to be observed.

\item  {\bf Halo bias :} Our merger rate predictions are biased since they are based on zoom-in simulations targeting massive DM halos. To quantify this bias, we adopted the semi-analytical model by \citet{Barausse_2012} and showed that a MBH model like ours (with heavy BH seeds and instantaneous mergers) overpredicts the actual merger rate by a factor of $\sim $ 20 at comparable redshifts (z > 6). For lighter seeds as well as heavy seed models with time delays included, the merger rate overprediction still remains at a factor between 20 to 30.

\item  {\bf Delays in MBHB mergers :} We corrected in post-processing the coalescing time of MBHB mergers including a time delay due to dynamical friction from the surrounding stars. We found that, if this delay is considered, 83 per cent  of MBHBs will merge within the Hubble time, but only 21 per cent of them, will merge within 1 Gyr, namely the age of the Universe at $z > 6$.

\item {\bf Signal-to-noise ratio and angular resolution :} Taking into account the LISA frequency bands, we calculated the signal-to-noise ratio and the angular resolution of the GW events predicted by our fiducial run, \AGNfiducial{}. The fraction of LISA detectable events  with high signal-to-noise ratio (SNR$>$5) ranges between 66-69 per cent depending on the inclusion of time delays in post-processing. The largest SNR is reached in the case of low chirp masses ($<10^6 \rm \msun$) which, although being characterized by a smaller characteristic strain, remain in the LISA band for a longer time, thus decreasing the noise and increasing the SNR. These systems are however, characterized by very low angular resolutions (10 $\rm deg^2$).

\item {\bf Mass ratio of LDEs :}  We computed the distribution of the mass ratio of the LDEs in the \AGNfiducial{} run and we found that the maximum number of mergers occurs for equi-massed binaries which are 'just-seeded' (i.e. with $M_{\rm BH} \sim 10^5 \msun$). For a fixed total mass of a MBHB, the number of mergers increases with increasing mass ratio which can be attributed to hierarchical structure formation: several low-mass MBHs merge (and accrete) to form more massive MBHs. This remained true even when further time delays in merging are considered.

\end{itemize}

One of the main goals of our study was to quantify the range of uncertainties on the merger rate of LISA detectable events. We find that considerable different merger rates result from the simulations in our suite. To get a more reliable constraint on this important issue, the shortage of currently adopted models should be addressed. 
For what concerns the EM signals arising from LDEs, the main challenge remains the poor LISA sky localization that in the most optimistic case ($\mathcal M_c\sim 10^6~\msun$ at $z=6$) is around 10~$\rm deg^2$. For this reason, it is important to further investigate the EM properties of LDEs to search for eventual, unique signatures from MBH coalescences to maximize the chances of their detections which we address in our future work.

\section*{Acknowledgements}
The authors thank Andrea Pallottini, Hamsa Padmanabhan and Alessandro Lupi for their helpful insights and comments. SG acknowledges support from the ASI-INAF n. 2018-31-HH.0 grant and PRIN-MIUR 2017. MV is supported by the Alexander von Humboldt Stiftung and the Carl Friedrich von Siemens Stiftung. MV also acknowledges support from the Excellence Cluster ORIGINS, which is funded by the Deutsche Forschungsgemeinschaft (DFG, German Research Foundation) under Germany's Excellence Strategy - EXC-2094 - 390783311. PB acknowledges support from the Brazilian funding agency FAPESP (grants 2016/01355-5, 2016/22183-8). D.I.V.  acknowledges the financial support provided under the European Union’s H2020 ERC Consolidator Grant ``Binary Massive Black Hole Astrophysics'' (B Massive, Grant Agreement: 818691)

\section*{Data Availability Statement}
The data underlying this article will be shared on reasonable request to the corresponding author.

\bibliographystyle{mnras}
\bibliography{references} % if your bibtex file is called example.bib

%%%%%%%%%%%%%%%%% APPENDICES %%%%%%%%%%%%%%%%%%%%%

%\setcounter{table}{0}
%\renewcommand{\thetable}{A\arabic{table}}
%\appendix
%\section{List of merger events in \AGNfiducial{}}

% \usepackage{graphicx}
\appendix

\section{Constraints from PTA}
\label{PTA}

We compare our predictions with upper limits placed by several Pulsar Timing Array (PTA) observations.
PTAs are sensitive to GWs with frequencies between $10^{-9}$ to $10^{-7}$ Hz. In this nano-Hz regime, the signal mostly arises from stochastic gravitational wave background (GWB) produced by the incoherent superposition of GWs from the population of inspiralling MBHBs overlapping in frequencies.
The characteristic strain arising from this stochastic GWB can be written as \citep{Sesana_2008_PTA}:
\begin{equation} \label{eq:PTA1}
    h^2_c(f) \, {=} \, \frac{4 G^{5/3}}{f^{2} c^2\pi} {\int}{\int} \frac{dz d\mathcal{M}}{(1+z)} \frac{d^2N}{dzd\mathcal{M}}\frac{d\mathrm{E}_{\rm GW}(\mathcal{M})}{d\ln{f_r}},
\end{equation}
where $d^2N/dzd\mathcal{M}$ is the comoving number density of MBHB merger per unit redshift, and rest-frame chirp mass, $f$ is the frequency of the GWs in the observer frame, and $d\mathrm{E_{GW}}/d\ln{f_r}$ is the energy emitted per logarithmic rest-frame frequency, $f_r$. 

Assuming the inspiralling population of MBHBs in the PTA band are in perfect circular orbits, Eq.~\ref{eq:PTA1} can be re-written as:
\begin{equation} \label{eq:PTA2}
    h^2_c(f) \, {=} \, \frac{4 G^{5/3} f^{-4/3}}{3c^2\pi^{1/3}} {\int}{\int} dz d\mathcal{M} \frac{d^2N}{dzd\mathcal{M}} \frac{\mathcal{M}^{5/3}}{(1+z)^{1/3}}.
\end{equation}
This type of relation is typically written as:
\begin{equation}\label{eq:PTA3}
h_c(f) \,{=}\, A\left( \frac{f}{f_0} \right)^{-2/3}, 
\end{equation}
where $A$ is the amplitude of the signal at the reference frequency $f_0$ which is usually normalized at $f_0 \,{=}\ \rm 1yr^{-1}$, and $\mathrm{A}(f_0\,{=}\,1 \rm yr^{-1})$ is typically denoted as $A_{\rm yr^{-1}}$. 
\begin{figure}
        \includegraphics[width=\columnwidth]{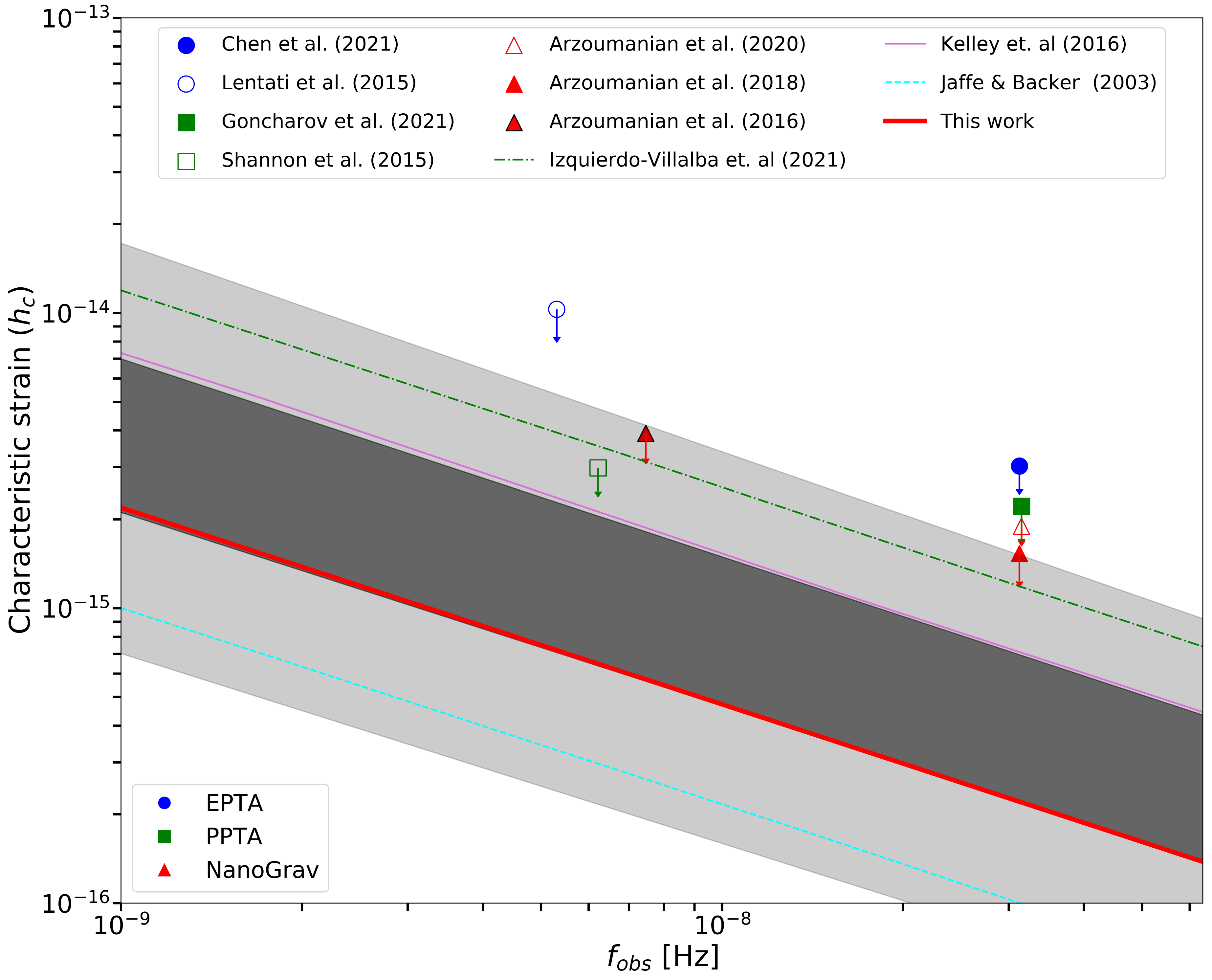}

        \caption{Stochastic Gravitational Wave Background spectrum calculated from our fiducial model (red solid line). The circles, squares, and triangles represent EPTA, PPTA and NanoGrav data respectively. Dark (light) gray shaded region depicts the $1\sigma$ ($2\sigma$) confidence level of the predictions by \citet{Sesana_2016}. The green dot-dashed line shows the results by \citet{Izquierdo_Villalba_2021}: $A_{\rm yr^{-1}}$  ${\sim}\,1.2{\times}10^{-15}$.  The solid pink line refers to the \texttt{Illustris} predictions of $A_{\rm yr^{-1}} \sim\,7.1{\times}10^{-16}$ \citep{Kelley_2016}. The cyan dashed line shows the results by \citet{Jaffe_2003}: $A_{\rm yr^{-1}}{\sim}\,1{\times}10^{-16}$.}
       \label{fig:GW_PTA} 
\end{figure}

We present our predictions in Fig.~\ref{fig:GW_PTA}, where we apply a bias correction factor of 20 (see Sec. \ref{sec:halobias}). We find $A_{\rm yr^{-1}}\sim\,2 {\times}10^{-16}$, which is below the upper limits placed by the NANOGrav \citep{Arzoumanian_2018}, EPTA \citep{Lentati_2015} observations and the PPTA observations \citep{Shannon_2015}. We emphasize that our predictions do not include any contribution from sources at $z<6$, which are instead expected to dominate the background \citep[e.g.][]{Izquierdo_Villalba_2021}.

\section{Time delay due to stellar hardening} \label{stellar hardening}

Stellar hardening is the process by which stars interact with the MBHB and gradually become more tightly bound to it over time. For already bound MBHBs, the gravitational field can be strong enough to disrupt the orbits of nearby stars. As a result, some of these stars can be captured by the MBHB and start to orbit around it \citep{Mikkola92}. As these stars continue to interact with the binary, they can extract energy and angular momentum from the MBHB, causing it to become more tightly bound \citep{Quinlan_1996}.

We compute the stellar hardening timescale following \citet{Sesana_2015}:
\begin{equation}
t_{\rm stellar}=15.18\, {\rm Gyr} \left(\frac{\sigma_{\rm inf}}{\kms}\right) \left(\frac{\rho_{\rm inf}}{\msun {\rm pc}^{-3}}\right)^{-1}\left(\frac{a_{\rm gw}}{10^{-3}\rm{pc}}\right)^{-1}, 
\label{eq:tstell}
\end{equation}
where $\sigma_{\rm inf}$ and $\rho_{\rm inf}$ are the velocity dispersion and the stellar density at the sphere of influence\footnote{The sphere of influence is approximated as the sphere containing twice the binary mass in stars.}, and $a_{\rm gw}$ is the transition separation of stellar hardening and GW hardening at which the binary spends most of its time: 
\begin{equation}
\sigma_{\rm inf}=(GM_{\rm t}/r_{\rm inf})^{1/2},
\label{eq:siginf}
\end{equation}
\begin{equation}
\rho_{\rm inf}=\frac{(3-\gamma)M_{*}r_{\rm inf}^{-\gamma}}{8\pi R_{\rm eff}^{3-\gamma}},
\label{eq:rhoinf}
\end{equation}
\begin{equation}
\begin{split}
a_{\rm gw} = & 2.64 \times 10^{-2} {\rm pc} \, \times \\
 & \left[\frac{\sigma_{\rm inf}}{\kms}\,\frac{\msun {\rm pc}^{-3}}{\rho_{\rm inf}}\,\frac{15}{H}\,\left (\frac{M_{\rm BH_p}\,M_{\rm BH_s}\,M_{t}}{2\times10^{24} \rm M^3_\odot}\right )\right]^{1/5},
\end{split}
\label{eq:agw}
\end{equation}
where $r_{\rm inf}$ is the radius containing twice the binary mass in stars:
\begin{equation}
r_{\rm inf}=R_{\rm eff}\left(\frac{4M_{t}}{M_{*}}\right)^{1/(3-\gamma)},
\label{eq:rinf}
\end{equation}
and we assume the index $\gamma$ = 2 \citep{Volonteri2020} and $H=15$ \citet{Sesana_2015}.

Finally, the total time of the MBHB merger including dynamical friction and stellar hardening is given by:
\begin{equation}
t_{\rm tot,stellar}= t_{\rm tot,df} + t_{\rm stellar}
\label{eq:tstelltot}
\end{equation}

According to the formalism reported above, we find that $t_{\rm stellar}$ varies in the range 1-10 Gyr, since central stellar densities are in the range 40-600 $\msun \rm pc^{-3}$. These $\rho_{\rm inf}$ values are not a fair representation, likely because the resolution of our simulations do not allow us to properly determine the matter distribution on such small scales. For comparison, in Sgr A$^*$ $\rho_{\rm inf}\sim 7 \times 10^4\msun \rm pc^{-3}$. In addition, galaxies at high redshift are likely more centrally concentrated than local galaxies. Thus, we would expect $\rho_{\rm inf}$ values even larger than what is found in Sgr A$^*$. For all these reasons, we do not include the delay due to stellar hardening in our post-processing calculations.

\section{GW detectables for different simulation runs} \label{GW_all}

Here we show the comparison of different GW detectables for the simulation runs \AGNcone{}, \AGNsphere{} and \BHsnoFB{} compared with \AGNfiducial{}. In all cases we report the results for a biased halo with no delays in post-processing.
\begin{figure*}
        \includegraphics[width=1.6\columnwidth]{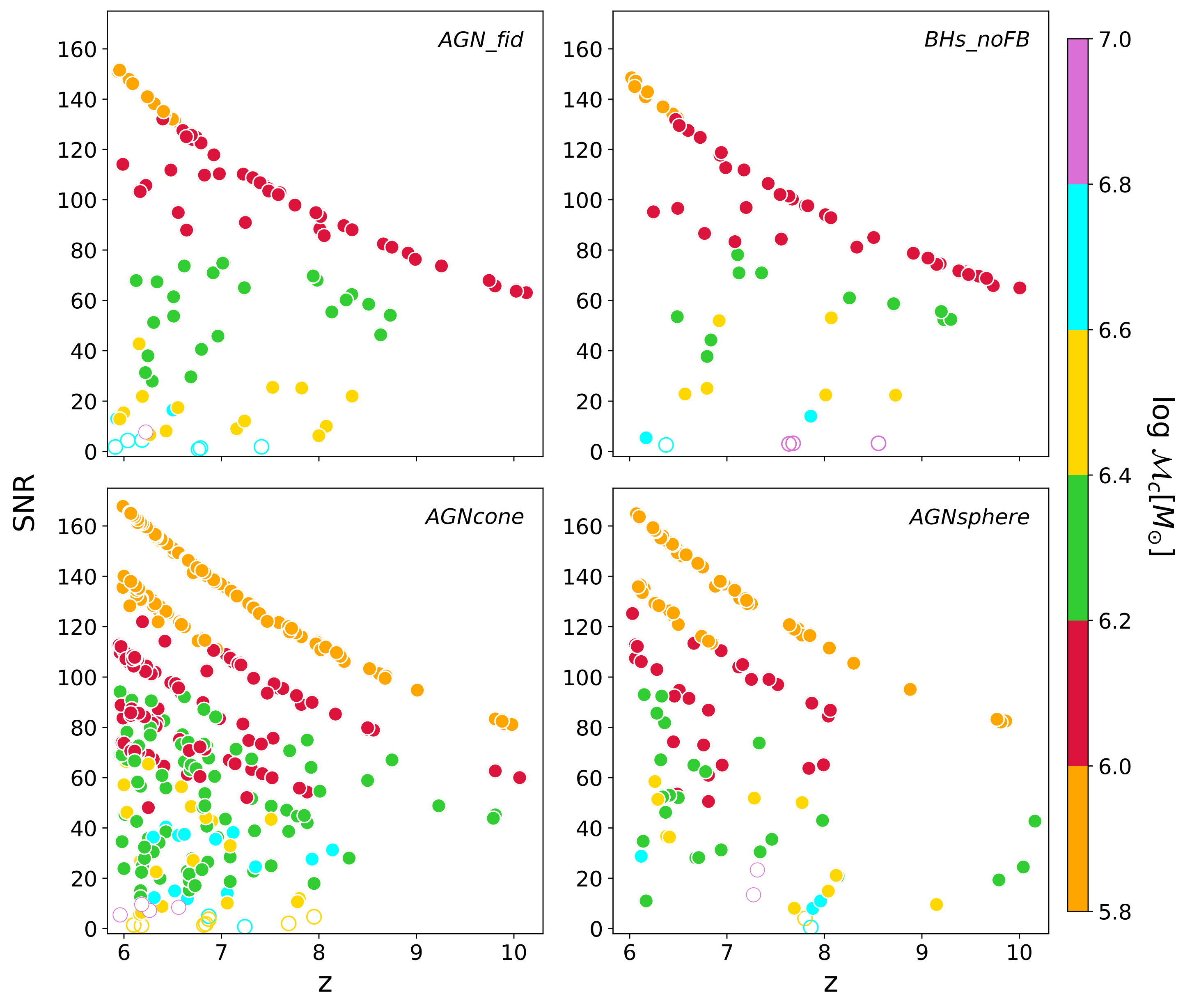}
	\includegraphics[width=1.6\columnwidth]{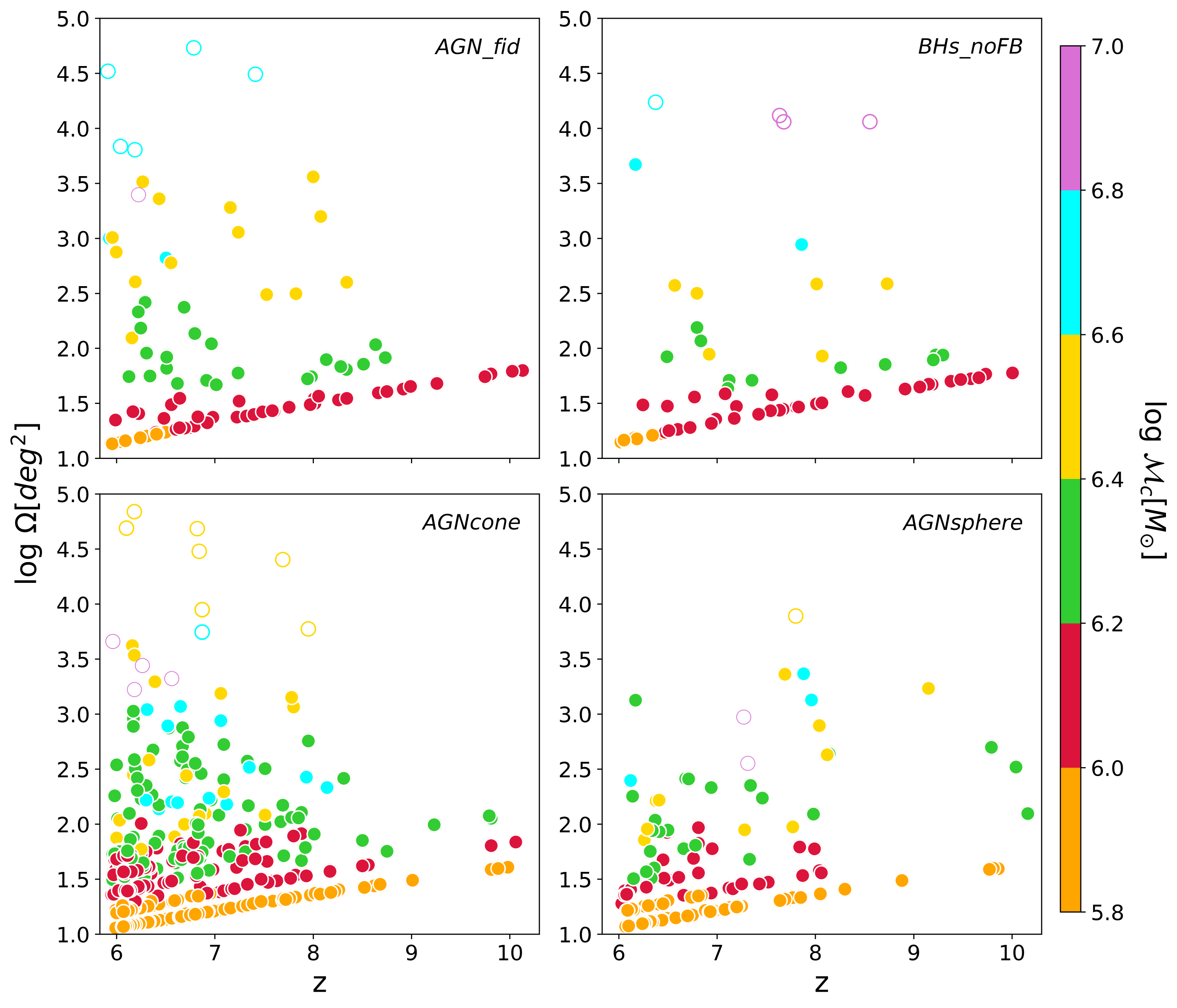}
    \caption{ Signal-to-noise ratio (upper panel) and angular resolution (lower panel) of MBHB mergers resulting from different runs of our simulations, color-coded according to their chirp mass.  Filled and empty circles represents detectable and undetectable events, respectively. The detectability threshold has been set to SNR$_{\rm thres}$ = 5.}
    \label{fig:snrold}
\end{figure*}
Fig. \ref{fig:snrold} shows the redshift distribution of SNR (upper panel) and angular resolution (lower panel) of the GW events for different chirp masses in our four different simulation suits. The overall trend in the SNR and resolution distribution is explained in Sec. \ref{snrtext}. Since only in the \citetalias{Barai2018} simulations the chirp mass PDF is populated for $\mathcal M_c<10^6\msun$, this explains why for these simulations the SNR predicted reaches the highest value (SNR$>$160). Apart from this aspect (and the different number of events predicted, already discussed in Sec. \ref{sec:mrates}), we do not find in the SNR predictions huge differences among different models. Consequently, the corresponding angular resolutions of event the loudest GW events remain quite poor ($\sim 10 deg^2$) for all models.

\begin{figure}
	
        \includegraphics[width=\columnwidth]{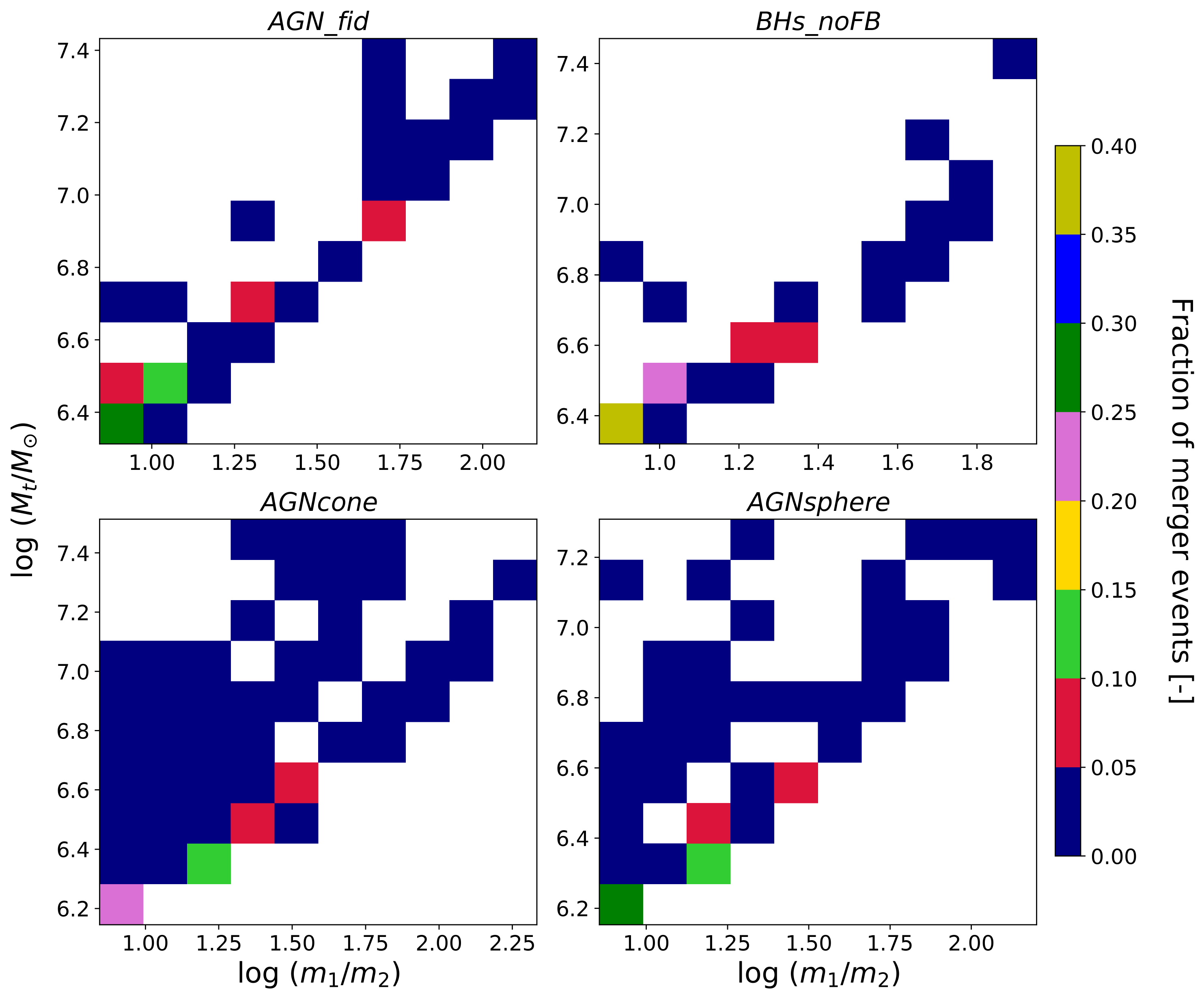}
    \caption{Fraction of detectable merger events as a function of the total mass of the system $M_t=m_1+m_2$ and the mass ratio $m_1/m_2$ of the binary.}
    \label{fig:massratioold}
\end{figure}

In Fig. \ref{fig:massratioold}, we also show the fraction of mergers are distributed in  the "mass ratio" and "total mass" plane, in the four simulation runs analysed in this work. The different coverage of the "mass ratio"-"total mass" plane simply reflects the different numbers of LDEs in the simulations runs: as shown in table \ref{lisatab} a higher number of LDEs occurs in the \AGNcone{} run, which shows the most densely populated "mass ratio"-"total mass" plane, while the smaller number of LDEs is predicted in the \BHsnoFB{} run.

As seen in Sec.\ref{reobs}, highest fraction of LDEs occurs for equi-mass binaries, in the low mass range ($M_t\lesssim 3 \times 10^{6} \msun$), independent of the feedback implemented and the simulation resolution and it ranges between 20 per cent and 40 per cent of the total LDE population in each run.

%%%%%%%%%%%%%%%%%%%%%%%%%%%%%%%%%%%%%%%%%%%%%%%%%%
% Don't change these lines
\bsp	% typesetting comment
\label{lastpage}

\end{document}

%% file: include_tex/definitions.tex
\def\be{\begin{equation}} 
\def\ee{\end{equation}} 
\def\ba{\begin{eqnarray}} 
\def\ea{\end{eqnarray}}

\def\kms{\,{\rm {km\, s^{-1}}}} 
\def\cc{\,{\rm {cm^{-3}}}} 
\def\msun{{\Msun}}

\def\HH{${\rm {H_2}}$}

\def\gsim{\lower.5ex\hbox{\gtsima}} 
\def\lsim{\lower.5ex\hbox{\ltsima}} \def\gtsima{$\; \buildrel > \over 
\sim \;$} \def\ltsima{$\; \buildrel < \over \sim \;$} \def\prosima{$\; 
\buildrel \propto \over \sim \;$} \def\gsim{\lower.5ex\hbox{\gtsima}} 
\def\lsim{\lower.5ex\hbox{\ltsima}} 
\def\simgt{\lower.5ex\hbox{\gtsima}} 
\def\simlt{\lower.5ex\hbox{\ltsima}} 
\def\simpr{\lower.5ex\hbox{\prosima}} \def\la{\lsim} \def\ga{\gsim} 
  
 \def\gtsima{$\; \buildrel > \over \sim \;$} 
\def\ltsima{$\; \buildrel < \over \sim \;$} 
\def\gsim{\lower.5ex\hbox{\gtsima}} 
\def\lsim{\lower.5ex\hbox{\ltsima}} 
\def\simgt{\lower.5ex\hbox{\gtsima}} 
\def\simlt{\lower.5ex\hbox{\ltsima}} 
\def\simpr{\lower.5ex\hbox{\prosima}} 
\def\la{\lsim} 
\def\ga{\gsim}

\def\msun{\,{\rm \Msun}}

\def\E3{{\cal E}_{\rm g}^{III}}

\def\r12{r_{1/2}} 
\def\x12{x_{1/2}} 
\def\v12{v_{1/2}}

%% file: include_tex/additional_def.tex
\newcommand\code[1]{\textsc{\MakeLowercase{#1}}}

\def\nh2{n_{\rm H2}}
\def\fh2{f_{\rm H2}}

\def\angstrom{\textrm{A\kern -1.3ex\raisebox{0.6ex}{$^\circ$}}}

\def\msun{{\rm M}_{\odot}}

\def\cc{{\rm cm}^{-3}}

\def\AGNsphere{\emph{AGNsphere}}
\def\AGNcone{\emph{AGNcone}}
\def\AGNfiducial{\emph{AGN\_fid}}
\def\BHsnoFB{\emph{BHs\_noFB}}

%% file: mnras_template.bbl
\begin{thebibliography}{}
\makeatletter
\relax
\def\mn@urlcharsother{\let\do\@makeother \do\$\do\&\do\#\do\^\do\_\do\%\do\~}
\def\mn@doi{\begingroup\mn@urlcharsother \@ifnextchar [ {\mn@doi@}
  {\mn@doi@[]}}
\def\mn@doi@[#1]#2{\def\@tempa{#1}\ifx\@tempa\@empty \href
  {http://dx.doi.org/#2} {doi:#2}\else \href {http://dx.doi.org/#2} {#1}\fi
  \endgroup}
\def\mn@eprint#1#2{\mn@eprint@#1:#2::\@nil}
\def\mn@eprint@arXiv#1{\href {http://arxiv.org/abs/#1} {{\tt arXiv:#1}}}
\def\mn@eprint@dblp#1{\href {http://dblp.uni-trier.de/rec/bibtex/#1.xml}
  {dblp:#1}}
\def\mn@eprint@#1:#2:#3:#4\@nil{\def\@tempa {#1}\def\@tempb {#2}\def\@tempc
  {#3}\ifx \@tempc \@empty \let \@tempc \@tempb \let \@tempb \@tempa \fi \ifx
  \@tempb \@empty \def\@tempb {arXiv}\fi \@ifundefined
  {mn@eprint@\@tempb}{\@tempb:\@tempc}{\expandafter \expandafter \csname
  mn@eprint@\@tempb\endcsname \expandafter{\@tempc}}}

\bibitem[\protect\citeauthoryear{{Amaro-Seoane} et~al.,}{{Amaro-Seoane}
  et~al.}{2022}]{2022arXiv220306016A}
{Amaro-Seoane} P.,  et~al., 2022, arXiv e-prints, \href
  {https://ui.adsabs.harvard.edu/abs/2022arXiv220306016A} {p. arXiv:2203.06016}

\bibitem[\protect\citeauthoryear{Arun et~al.,}{Arun et~al.}{2009}]{Arun_2009}
Arun K.~G.,  et~al., 2009, \mn@doi [Classical and Quantum Gravity]
  {10.1088/0264-9381/26/9/094027}, 26, 094027

\bibitem[\protect\citeauthoryear{Arzoumanian et~al.,}{Arzoumanian
  et~al.}{2018}]{Arzoumanian_2018}
Arzoumanian Z.,  et~al., 2018, \mn@doi [The Astrophysical Journal]
  {10.3847/1538-4357/aabd3b}, 859, 47

\bibitem[\protect\citeauthoryear{Barai \& de Gouveia~Dal~Pino}{Barai \&
  de~Gouveia~Dal~Pino}{2019}]{Barai_2019}
Barai P.,  de Gouveia~Dal~Pino E.~M.,  2019, \mn@doi [Monthly Notices of the
  Royal Astronomical Society] {10.1093/mnras/stz1616}, 487, 5549

\bibitem[\protect\citeauthoryear{{Barai}, {Gallerani}, {Pallottini}, {Ferrara},
  {Marconi}, {Cicone}, {Maiolino}  \& {Carniani}}{{Barai}
  et~al.}{2018}]{Barai2018}
{Barai} P.,  {Gallerani} S.,  {Pallottini} A.,  {Ferrara} A.,  {Marconi} A.,
  {Cicone} C.,  {Maiolino} R.,   {Carniani} S.,  2018, \mn@doi [\mnras]
  {10.1093/mnras/stx2563}, \href
  {https://ui.adsabs.harvard.edu/abs/2018MNRAS.473.4003B} {473, 4003}

\bibitem[\protect\citeauthoryear{Barausse}{Barausse}{2012}]{Barausse_2012}
Barausse E.,  2012, \mn@doi [Monthly Notices of the Royal Astronomical Society]
  {10.1111/j.1365-2966.2012.21057.x}, 423, 2533

\bibitem[\protect\citeauthoryear{Barausse, Dvorkin, Tremmel, Volonteri  \&
  Bonetti}{Barausse et~al.}{2020}]{Barausse_2020}
Barausse E.,  Dvorkin I.,  Tremmel M.,  Volonteri M.,   Bonetti M.,  2020,
  \mn@doi [The Astrophysical Journal] {10.3847/1538-4357/abba7f}, 904, 16

\bibitem[\protect\citeauthoryear{{Barkana} \& {Loeb}}{{Barkana} \&
  {Loeb}}{2001}]{Barkana_Loeb2001}
{Barkana} R.,  {Loeb} A.,  2001, \mn@doi [\physrep]
  {10.1016/S0370-1573(01)00019-9}, \href
  {https://ui.adsabs.harvard.edu/abs/2001PhR...349..125B} {349, 125}

\bibitem[\protect\citeauthoryear{Begelman, Volonteri  \& Rees}{Begelman
  et~al.}{2006}]{Begelman_2006}
Begelman M.~C.,  Volonteri M.,   Rees M.~J.,  2006, \mn@doi [Monthly Notices of
  the Royal Astronomical Society] {10.1111/j.1365-2966.2006.10467.x}, 370,
  289–298

\bibitem[\protect\citeauthoryear{Bhowmick et~al.,}{Bhowmick
  et~al.}{2022}]{Bhowmick}
Bhowmick A.~K.,  et~al., 2022, Probing the $z\gtrsim6$ quasars in a universe
  with IllustrisTNG physics: Impact of gas-based black hole seeding models,
  \mn@doi{10.48550/ARXIV.2205.05717}, \url {https://arxiv.org/abs/2205.05717}

\bibitem[\protect\citeauthoryear{{Binney} \& {Tremaine}}{{Binney} \&
  {Tremaine}}{2008}]{Binney2008}
{Binney} J.,  {Tremaine} S.,  2008, {Galactic Dynamics: Second Edition}

\bibitem[\protect\citeauthoryear{Blanchet, Damour, Iyer, Will  \&
  Wiseman}{Blanchet et~al.}{1995}]{Blanchet_1995}
Blanchet L.,  Damour T.,  Iyer B.~R.,  Will C.~M.,   Wiseman A.~G.,  1995,
  \mn@doi [Physical Review Letters] {10.1103/physrevlett.74.3515}, 74, 3515

\bibitem[\protect\citeauthoryear{Blecha et~al.,}{Blecha
  et~al.}{2015}]{Blecha_2015}
Blecha L.,  et~al., 2015, \mn@doi [Monthly Notices of the Royal Astronomical
  Society] {10.1093/mnras/stv2646}, 456, 961

\bibitem[\protect\citeauthoryear{Blumenthal, Faber, Primack  \&
  Rees}{Blumenthal et~al.}{1984}]{Blumenthal:1984bp}
Blumenthal G.~R.,  Faber S.,  Primack J.~R.,   Rees M.~J.,  1984, \mn@doi
  [Nature] {10.1038/311517a0}, 311, 517

\bibitem[\protect\citeauthoryear{{Bondi}}{{Bondi}}{1952}]{Bondi52}
{Bondi} H.,  1952, \mn@doi [\mnras] {10.1093/mnras/112.2.195}, \href
  {https://ui.adsabs.harvard.edu/abs/1952MNRAS.112..195B} {112, 195}

\bibitem[\protect\citeauthoryear{{Bondi} \& {Hoyle}}{{Bondi} \&
  {Hoyle}}{1944}]{Bondi_Hoyle}
{Bondi} H.,  {Hoyle} F.,  1944, \mn@doi [\mnras] {10.1093/mnras/104.5.273},
  \href {https://ui.adsabs.harvard.edu/abs/1944MNRAS.104..273B} {104, 273}

\bibitem[\protect\citeauthoryear{Bonetti, Sesana, Haardt, Barausse  \&
  Colpi}{Bonetti et~al.}{2019}]{Bonetti}
Bonetti M.,  Sesana A.,  Haardt F.,  Barausse E.,   Colpi M.,  2019, \mn@doi
  [Monthly Notices of the Royal Astronomical Society] {10.1093/mnras/stz903},
  486, 4044–4060

\bibitem[\protect\citeauthoryear{Booth \& Schaye}{Booth \&
  Schaye}{2009}]{Booth_2009}
Booth C.~M.,  Schaye J.,  2009, \mn@doi [Monthly Notices of the Royal
  Astronomical Society] {10.1111/j.1365-2966.2009.15043.x}, 398, 53–74

\bibitem[\protect\citeauthoryear{Chabrier}{Chabrier}{2003}]{Chabrier_2003}
Chabrier G.,  2003, \mn@doi [Publications of the Astronomical Society of the
  Pacific] {10.1086/376392}, 115, 763–795

\bibitem[\protect\citeauthoryear{{Chandrasekhar}}{{Chandrasekhar}}{1943}]{Chandrasekhar1943}
{Chandrasekhar} S.,  1943, \mn@doi [\apj] {10.1086/144517}, \href
  {https://ui.adsabs.harvard.edu/abs/1943ApJ....97..255C} {97, 255}

\bibitem[\protect\citeauthoryear{Ciardi \& Loeb}{Ciardi \&
  Loeb}{2000}]{Ciardi2000}
Ciardi B.,  Loeb A.,  2000, \mn@doi [The Astrophysical Journal]
  {10.1086/309384}, 540, 687–696

\bibitem[\protect\citeauthoryear{Correa, Wyithe, Schaye  \& Duffy}{Correa
  et~al.}{2015a}]{Correa_2015}
Correa C.~A.,  Wyithe J. S.~B.,  Schaye J.,   Duffy A.~R.,  2015a, \mn@doi
  [Monthly Notices of the Royal Astronomical Society] {10.1093/mnras/stv689},
  450, 1514

\bibitem[\protect\citeauthoryear{Correa, Wyithe, Schaye  \& Duffy}{Correa
  et~al.}{2015b}]{Correa_2015b}
Correa C.~A.,  Wyithe J. S.~B.,  Schaye J.,   Duffy A.~R.,  2015b, \mn@doi
  [Monthly Notices of the Royal Astronomical Society] {10.1093/mnras/stv697},
  450, 1521

\bibitem[\protect\citeauthoryear{Cutler \& Flanagan}{Cutler \&
  Flanagan}{1994}]{Cutler_1994}
Cutler C.,  Flanagan {\'{E} }.~E.,  1994, \mn@doi [Physical Review D]
  {10.1103/physrevd.49.2658}, 49, 2658

\bibitem[\protect\citeauthoryear{{Davies}, {Miller}  \& {Bellovary}}{{Davies}
  et~al.}{2011}]{Davies2011ApJ...740L..42D}
{Davies} M.~B.,  {Miller} M.~C.,   {Bellovary} J.~M.,  2011, \mn@doi [\apjl]
  {10.1088/2041-8205/740/2/L42}, \href
  {https://ui.adsabs.harvard.edu/abs/2011ApJ...740L..42D} {740, L42}

\bibitem[\protect\citeauthoryear{Dayal, Rossi, Shiralilou, Piana, Choudhury  \&
  Volonteri}{Dayal et~al.}{2019}]{Dayal2019}
Dayal P.,  Rossi E.~M.,  Shiralilou B.,  Piana O.,  Choudhury T.~R.,
  Volonteri M.,  2019, \mn@doi [Monthly Notices of the Royal Astronomical
  Society] {10.1093/mnras/stz897}, 486, 2336–2350

\bibitem[\protect\citeauthoryear{DeGraf \& Sijacki}{DeGraf \&
  Sijacki}{2019}]{DeGraf2020}
DeGraf C.,  Sijacki D.,  2019, \mn@doi [Monthly Notices of the Royal
  Astronomical Society] {10.1093/mnras/stz3309}, 491, 4973–4992

\bibitem[\protect\citeauthoryear{{DeGraf}, {Sijacki}, {Di Matteo},
  {Holley-Bockelmann}, {Snyder}  \& {Springel}}{{DeGraf}
  et~al.}{2021}]{DeGraf2021}
{DeGraf} C.,  {Sijacki} D.,  {Di Matteo} T.,  {Holley-Bockelmann} K.,  {Snyder}
  G.,   {Springel} V.,  2021, \mn@doi [\mnras] {10.1093/mnras/stab721}, \href
  {https://ui.adsabs.harvard.edu/abs/2021MNRAS.503.3629D} {503, 3629}

\bibitem[\protect\citeauthoryear{{Devecchi}, {Volonteri}, {Rossi}, {Colpi}  \&
  {Portegies Zwart}}{{Devecchi} et~al.}{2012}]{Devecchi2012MNRAS.421.1465D}
{Devecchi} B.,  {Volonteri} M.,  {Rossi} E.~M.,  {Colpi} M.,   {Portegies
  Zwart} S.,  2012, \mn@doi [\mnras] {10.1111/j.1365-2966.2012.20406.x}, \href
  {https://ui.adsabs.harvard.edu/abs/2012MNRAS.421.1465D} {421, 1465}

\bibitem[\protect\citeauthoryear{Di~Matteo, Khandai, DeGraf, Feng, Croft, Lopez
   \& Springel}{Di~Matteo et~al.}{2012}]{Di_Matteo_2012}
Di~Matteo T.,  Khandai N.,  DeGraf C.,  Feng Y.,  Croft R. A.~C.,  Lopez J.,
  Springel V.,  2012, \mn@doi [The Astrophysical Journal]
  {10.1088/2041-8205/745/2/l29}, 745, L29

\bibitem[\protect\citeauthoryear{Edgar}{Edgar}{2004}]{Edgar_2004}
Edgar R.,  2004, \mn@doi [New Astronomy Reviews] {10.1016/j.newar.2004.06.001},
  48, 843–859

\bibitem[\protect\citeauthoryear{Eisenstein \& Loeb}{Eisenstein \&
  Loeb}{1995}]{Eisenstein_1995}
Eisenstein D.~J.,  Loeb A.,  1995, \mn@doi [The Astrophysical Journal]
  {10.1086/175498}, 443, 11

\bibitem[\protect\citeauthoryear{Fan et~al.,}{Fan et~al.}{2006}]{Fan_2006}
Fan X.,  et~al., 2006, \mn@doi [The Astronomical Journal] {10.1086/504836},
  132, 117–136

\bibitem[\protect\citeauthoryear{Farmer \& Phinney}{Farmer \&
  Phinney}{2003}]{Farmer_2003}
Farmer A.~J.,  Phinney E.~S.,  2003, \mn@doi [Monthly Notices of the Royal
  Astronomical Society] {10.1111/j.1365-2966.2003.07176.x}, 346, 1197–1214

\bibitem[\protect\citeauthoryear{Ferrara, Salvadori, Yue  \&
  Schleicher}{Ferrara et~al.}{2014}]{Ferrara_2014}
Ferrara A.,  Salvadori S.,  Yue B.,   Schleicher D.,  2014, \mn@doi [Monthly
  Notices of the Royal Astronomical Society] {10.1093/mnras/stu1280}, 443,
  2410–2425

\bibitem[\protect\citeauthoryear{{Flanagan} \& {Hughes}}{{Flanagan} \&
  {Hughes}}{1998}]{Flanagan_1998}
{Flanagan} {\'E}.~{\'E}.,  {Hughes} S.~A.,  1998, \mn@doi [\prd]
  {10.1103/PhysRevD.57.4535}, \href
  {https://ui.adsabs.harvard.edu/abs/1998PhRvD..57.4535F} {57, 4535}

\bibitem[\protect\citeauthoryear{Haehnelt}{Haehnelt}{1994}]{Haehnelt_1994}
Haehnelt M.~G.,  1994, \mn@doi [Monthly Notices of the Royal Astronomical
  Society] {10.1093/mnras/269.1.199}, 269, 199–208

\bibitem[\protect\citeauthoryear{Hahn \& Abel}{Hahn \& Abel}{2011}]{Hahn_2011}
Hahn O.,  Abel T.,  2011, \mn@doi [Monthly Notices of the Royal Astronomical
  Society] {10.1111/j.1365-2966.2011.18820.x}, 415, 2101–2121

\bibitem[\protect\citeauthoryear{Hartwig, Agarwal  \& Regan}{Hartwig
  et~al.}{2018}]{Hartwig_2018}
Hartwig T.,  Agarwal B.,   Regan J.~A.,  2018, \mn@doi [Monthly Notices of the
  Royal Astronomical Society: Letters] {10.1093/mnrasl/sly091}, 479, L23–L27

\bibitem[\protect\citeauthoryear{{Hawking} \& {Israel}}{{Hawking} \&
  {Israel}}{1989}]{Hawking1989}
{Hawking} S.~W.,  {Israel} W.,  1989, {Three Hundred Years of Gravitation}.
Cambridge University Press

\bibitem[\protect\citeauthoryear{Heger, Fryer, Woosley, Langer  \&
  Hartmann}{Heger et~al.}{2003}]{Heger_2003}
Heger A.,  Fryer C.~L.,  Woosley S.~E.,  Langer N.,   Hartmann D.~H.,  2003,
  \mn@doi [The Astrophysical Journal] {10.1086/375341}, 591, 288–300

\bibitem[\protect\citeauthoryear{Hirano, Hosokawa, Yoshida, Omukai  \&
  Yorke}{Hirano et~al.}{2015}]{Hirano_2015}
Hirano S.,  Hosokawa T.,  Yoshida N.,  Omukai K.,   Yorke H.~W.,  2015, \mn@doi
  [Monthly Notices of the Royal Astronomical Society] {10.1093/mnras/stv044},
  448, 568

\bibitem[\protect\citeauthoryear{{Hoyle} \& {Lyttleton}}{{Hoyle} \&
  {Lyttleton}}{1939}]{Hoyle1939}
{Hoyle} F.,  {Lyttleton} R.~A.,  1939, \mn@doi [Proceedings of the Cambridge
  Philosophical Society] {10.1017/S0305004100021150}, \href
  {https://ui.adsabs.harvard.edu/abs/1939PCPS...35..405H} {35, 405}

\bibitem[\protect\citeauthoryear{Izquierdo-Villalba, Sesana, Bonoli  \&
  Colpi}{Izquierdo-Villalba et~al.}{2021}]{Izquierdo_Villalba_2021}
Izquierdo-Villalba D.,  Sesana A.,  Bonoli S.,   Colpi M.,  2021, \mn@doi
  [Monthly Notices of the Royal Astronomical Society] {10.1093/mnras/stab3239},
  509, 3488

\bibitem[\protect\citeauthoryear{Jaffe \& Backer}{Jaffe \&
  Backer}{2003}]{Jaffe_2003}
Jaffe A.~H.,  Backer D.~C.,  2003, \mn@doi [The Astrophysical Journal]
  {10.1086/345443}, 583, 616–631

\bibitem[\protect\citeauthoryear{Jiang, Jing  \& Lin}{Jiang
  et~al.}{2010}]{Jiang_2010}
Jiang C.~Y.,  Jing Y.~P.,   Lin W.~P.,  2010, \mn@doi [Astronomy and
  Astrophysics] {10.1051/0004-6361/200913257}, 510, A60

\bibitem[\protect\citeauthoryear{Johnson, Whalen, Fryer  \& Li}{Johnson
  et~al.}{2012}]{Johnson_2012}
Johnson J.~L.,  Whalen D.~J.,  Fryer C.~L.,   Li H.,  2012, \mn@doi [The
  Astrophysical Journal] {10.1088/0004-637x/750/1/66}, 750, 66

\bibitem[\protect\citeauthoryear{{Katz} \& {Larson}}{{Katz} \&
  {Larson}}{2019}]{katz2019_detectability}
{Katz} M.~L.,  {Larson} S.~L.,  2019, \mn@doi [\mnras] {10.1093/mnras/sty3321},
  \href {https://ui.adsabs.harvard.edu/abs/2019MNRAS.483.3108K} {483, 3108}

\bibitem[\protect\citeauthoryear{Katz, Kelley, Dosopoulou, Berry, Blecha  \&
  Larson}{Katz et~al.}{2019}]{Katz_2019}
Katz M.~L.,  Kelley L.~Z.,  Dosopoulou F.,  Berry S.,  Blecha L.,   Larson
  S.~L.,  2019, \mn@doi [Monthly Notices of the Royal Astronomical Society]
  {10.1093/mnras/stz3102}

\bibitem[\protect\citeauthoryear{Kelley, Blecha  \& Hernquist}{Kelley
  et~al.}{2016}]{Kelley_2016}
Kelley L.~Z.,  Blecha L.,   Hernquist L.,  2016, \mn@doi [Monthly Notices of
  the Royal Astronomical Society] {10.1093/mnras/stw2452}, 464, 3131

\bibitem[\protect\citeauthoryear{Klein et~al.,}{Klein
  et~al.}{2016}]{Klein_2016}
Klein A.,  et~al., 2016, \mn@doi [Physical Review D]
  {10.1103/physrevd.93.024003}, 93

\bibitem[\protect\citeauthoryear{Knollmann \& Knebe}{Knollmann \&
  Knebe}{2009}]{Knollmann2009}
Knollmann S.~R.,  Knebe A.,  2009, \mn@doi [The Astrophysical Journal
  Supplement Series] {10.1088/0067-0049/182/2/608}, 182, 608–624

\bibitem[\protect\citeauthoryear{Kormendy}{Kormendy}{2001}]{Kormendy_2001}
Kormendy J.,  2001, \mn@doi [AIP Conference Proceedings] {10.1063/1.1419581}

\bibitem[\protect\citeauthoryear{Koushiappas, Bullock  \& Dekel}{Koushiappas
  et~al.}{2004}]{Koushiappas_2004}
Koushiappas S.~M.,  Bullock J.~S.,   Dekel A.,  2004, \mn@doi [Monthly Notices
  of the Royal Astronomical Society] {10.1111/j.1365-2966.2004.08190.x}, 354,
  292–304

\bibitem[\protect\citeauthoryear{Krolik, Volonteri, Dubois  \&
  Devriendt}{Krolik et~al.}{2019}]{Krolik_2019}
Krolik J.~H.,  Volonteri M.,  Dubois Y.,   Devriendt J.,  2019, \mn@doi [The
  Astrophysical Journal] {10.3847/1538-4357/ab24c9}, 879, 110

\bibitem[\protect\citeauthoryear{Latif, Schleicher, Schmidt  \& Niemeyer}{Latif
  et~al.}{2013}]{Latif_2013}
Latif M.~A.,  Schleicher D. R.~G.,  Schmidt W.,   Niemeyer J.,  2013, \mn@doi
  [Monthly Notices of the Royal Astronomical Society] {10.1093/mnras/stt834},
  433, 1607

\bibitem[\protect\citeauthoryear{Lentati et~al.,}{Lentati
  et~al.}{2015}]{Lentati_2015}
Lentati L.,  et~al., 2015, \mn@doi [Monthly Notices of the Royal Astronomical
  Society] {10.1093/mnras/stv1538}, 453, 2577

\bibitem[\protect\citeauthoryear{Lodato \& Natarajan}{Lodato \&
  Natarajan}{2006}]{Lodato_2006}
Lodato G.,  Natarajan P.,  2006, \mn@doi [Monthly Notices of the Royal
  Astronomical Society] {10.1111/j.1365-2966.2006.10801.x}, 371, 1813–1823

\bibitem[\protect\citeauthoryear{Loeb}{Loeb}{2016}]{Loeb_2016}
Loeb A.,  2016, \mn@doi [The Astrophysical Journal]
  {10.3847/2041-8205/819/2/l21}, 819, L21

\bibitem[\protect\citeauthoryear{Loeb \& Rasio}{Loeb \&
  Rasio}{1994}]{Loeb_1994}
Loeb A.,  Rasio F.~A.,  1994, \mn@doi [The Astrophysical Journal]
  {10.1086/174548}, 432, 52

\bibitem[\protect\citeauthoryear{Lupi, Colpi, Devecchi, Galanti  \&
  Volonteri}{Lupi et~al.}{2014}]{Lupi_2014}
Lupi A.,  Colpi M.,  Devecchi B.,  Galanti G.,   Volonteri M.,  2014, \mn@doi
  [Monthly Notices of the Royal Astronomical Society] {10.1093/mnras/stu1120},
  442, 3616–3626

\bibitem[\protect\citeauthoryear{{Lupi}, {Volonteri}, {Decarli}, {Bovino},
  {Silk}  \& {Bergeron}}{{Lupi} et~al.}{2019}]{Lupi2019}
{Lupi} A.,  {Volonteri} M.,  {Decarli} R.,  {Bovino} S.,  {Silk} J.,
  {Bergeron} J.,  2019, \mn@doi [\mnras] {10.1093/mnras/stz1959}, \href
  {https://ui.adsabs.harvard.edu/abs/2019MNRAS.488.4004L} {488, 4004}

\bibitem[\protect\citeauthoryear{{Madau}, {Pozzetti}  \& {Dickinson}}{{Madau}
  et~al.}{1998}]{Madau1998ApJ...498..106M}
{Madau} P.,  {Pozzetti} L.,   {Dickinson} M.,  1998, \mn@doi [\apj]
  {10.1086/305523}, \href
  {https://ui.adsabs.harvard.edu/abs/1998ApJ...498..106M} {498, 106}

\bibitem[\protect\citeauthoryear{Magorrian et~al.,}{Magorrian
  et~al.}{1998}]{Magorrian_1998}
Magorrian J.,  et~al., 1998, \mn@doi [The Astronomical Journal]
  {10.1086/300353}, 115, 2285–2305

\bibitem[\protect\citeauthoryear{{Mapelli}}{{Mapelli}}{2016}]{Mapelli2016MNRAS.459.3432M}
{Mapelli} M.,  2016, \mn@doi [\mnras] {10.1093/mnras/stw869}, \href
  {https://ui.adsabs.harvard.edu/abs/2016MNRAS.459.3432M} {459, 3432}

\bibitem[\protect\citeauthoryear{McGee, Sesana  \& Vecchio}{McGee
  et~al.}{2020}]{mcgee2020linking}
McGee S.,  Sesana A.,   Vecchio A.,  2020, Linking gravitational waves and
  X-ray phenomena with joint LISA and Athena observations (\mn@eprint {arXiv}
  {1811.00050})

\bibitem[\protect\citeauthoryear{McWilliams, Lang, Baker  \& Thorpe}{McWilliams
  et~al.}{2011}]{McWilliams_2011}
McWilliams S.~T.,  Lang R.~N.,  Baker J.~G.,   Thorpe J.~I.,  2011, \mn@doi
  [Physical Review D] {10.1103/physrevd.84.064003}, 84

\bibitem[\protect\citeauthoryear{Mikkola \& Valtonen}{Mikkola \&
  Valtonen}{1992}]{Mikkola92}
Mikkola S.,  Valtonen M.~J.,  1992, \mn@doi [Monthly Notices of the Royal
  Astronomical Society] {10.1093/mnras/259.1.115}, 259, 115

\bibitem[\protect\citeauthoryear{{Murante}, {Monaco}, {Giovalli}, {Borgani}  \&
  {Diaferio}}{{Murante} et~al.}{2010}]{Murante2010}
{Murante} G.,  {Monaco} P.,  {Giovalli} M.,  {Borgani} S.,   {Diaferio} A.,
  2010, \mn@doi [\mnras] {10.1111/j.1365-2966.2010.16567.x}, \href
  {https://ui.adsabs.harvard.edu/abs/2010MNRAS.405.1491M} {405, 1491}

\bibitem[\protect\citeauthoryear{{Murante}, {Monaco}, {Borgani}, {Tornatore},
  {Dolag}  \& {Goz}}{{Murante} et~al.}{2015}]{Murante2015}
{Murante} G.,  {Monaco} P.,  {Borgani} S.,  {Tornatore} L.,  {Dolag} K.,
  {Goz} D.,  2015, \mn@doi [\mnras] {10.1093/mnras/stu2400}, \href
  {https://ui.adsabs.harvard.edu/abs/2015MNRAS.447..178M} {447, 178}

\bibitem[\protect\citeauthoryear{Navarro, Frenk  \& White}{Navarro
  et~al.}{1996}]{Navarro1996}
Navarro J.~F.,  Frenk C.~S.,   White S. D.~M.,  1996, \mn@doi [The
  Astrophysical Journal] {10.1086/177173}, 462, 563

\bibitem[\protect\citeauthoryear{{Nelemans}, {Yungelson}  \& {Portegies
  Zwart}}{{Nelemans} et~al.}{2001}]{Nelemans}
{Nelemans} G.,  {Yungelson} L.~R.,   {Portegies Zwart} S.~F.,  2001, \mn@doi
  [\aap] {10.1051/0004-6361:20010683}, \href
  {https://ui.adsabs.harvard.edu/abs/2001A&A...375..890N} {375, 890}

\bibitem[\protect\citeauthoryear{{Nelson} et~al.,}{{Nelson}
  et~al.}{2015}]{nelson2015}
{Nelson} D.,  et~al., 2015, \mn@doi [Astronomy and Computing]
  {10.1016/j.ascom.2015.09.003}, \href
  {https://ui.adsabs.harvard.edu/abs/2015A&C....13...12N} {13, 12}

\bibitem[\protect\citeauthoryear{Ostriker}{Ostriker}{1999}]{Ostriker_1999}
Ostriker E.~C.,  1999, \mn@doi [The Astrophysical Journal] {10.1086/306858},
  513, 252

\bibitem[\protect\citeauthoryear{{Peebles}}{{Peebles}}{1980}]{Peebles}
{Peebles} P.~J.~E.,  1980, {The large-scale structure of the universe}.
Princeton University Press

\bibitem[\protect\citeauthoryear{{Pillepich} et~al.,}{{Pillepich}
  et~al.}{2018}]{Pillepich2018}
{Pillepich} A.,  et~al., 2018, \mn@doi [\mnras] {10.1093/mnras/stx2656}, \href
  {https://ui.adsabs.harvard.edu/abs/2018MNRAS.473.4077P} {473, 4077}

\bibitem[\protect\citeauthoryear{{Planck Collaboration} Ade et~al.,}{{Planck
  Collaboration} et~al.}{2016}]{Planck2016}
{Planck Collaboration} Ade P. A.~R.,  et~al., 2016, \mn@doi [Astronomy &
  Astrophysics] {10.1051/0004-6361/201525830}, 594, A13

\bibitem[\protect\citeauthoryear{{Press} \& {Schechter}}{{Press} \&
  {Schechter}}{1974}]{PressSchechter_1974}
{Press} W.~H.,  {Schechter} P.,  1974, \mn@doi [\apj] {10.1086/152650}, \href
  {https://ui.adsabs.harvard.edu/abs/1974ApJ...187..425P} {187, 425}

\bibitem[\protect\citeauthoryear{Quinlan}{Quinlan}{1996}]{Quinlan_1996}
Quinlan G.~D.,  1996, \mn@doi [New Astronomy] {10.1016/s1384-1076(96)00003-6},
  1, 35

\bibitem[\protect\citeauthoryear{Reines, Greene  \& Geha}{Reines
  et~al.}{2013}]{Reines2013}
Reines A.~E.,  Greene J.~E.,   Geha M.,  2013, \mn@doi [The Astrophysical
  Journal] {10.1088/0004-637x/775/2/116}, 775, 116

\bibitem[\protect\citeauthoryear{{Reinoso}, {Schleicher}, {Fellhauer},
  {Klessen}  \& {Boekholt}}{{Reinoso}
  et~al.}{2018}]{Reinoso2018A&A...614A..14R}
{Reinoso} B.,  {Schleicher} D.~R.~G.,  {Fellhauer} M.,  {Klessen} R.~S.,
  {Boekholt} T.~C.~N.,  2018, \mn@doi [\aap] {10.1051/0004-6361/201732224},
  \href {https://ui.adsabs.harvard.edu/abs/2018A&A...614A..14R} {614, A14}

\bibitem[\protect\citeauthoryear{Salcido, Bower, Theuns, McAlpine, Schaller,
  Crain, Schaye  \& Regan}{Salcido et~al.}{2016}]{Salcido}
Salcido J.,  Bower R.~G.,  Theuns T.,  McAlpine S.,  Schaller M.,  Crain R.~A.,
   Schaye J.,   Regan J.,  2016, \mn@doi [Monthly Notices of the Royal
  Astronomical Society] {10.1093/mnras/stw2048}, 463, 870

\bibitem[\protect\citeauthoryear{{Schaye} et~al.,}{{Schaye}
  et~al.}{2015}]{Schaye2015MNRAS}
{Schaye} J.,  et~al., 2015, \mn@doi [\mnras] {10.1093/mnras/stu2058}, \href
  {https://ui.adsabs.harvard.edu/abs/2015MNRAS.446..521S} {446, 521}

\bibitem[\protect\citeauthoryear{Sesana \& Khan}{Sesana \&
  Khan}{2015}]{Sesana_2015}
Sesana A.,  Khan F.~M.,  2015, \mn@doi [Monthly Notices of the Royal
  Astronomical Society: Letters] {10.1093/mnrasl/slv131}, 454, L66

\bibitem[\protect\citeauthoryear{Sesana, Haardt, Madau  \& Volonteri}{Sesana
  et~al.}{2004}]{Sesana_2004}
Sesana A.,  Haardt F.,  Madau P.,   Volonteri M.,  2004, \mn@doi [The
  Astrophysical Journal] {10.1086/422185}, 611, 623–632

\bibitem[\protect\citeauthoryear{Sesana, Haardt, Madau  \& Volonteri}{Sesana
  et~al.}{2005}]{Sesana_2005}
Sesana A.,  Haardt F.,  Madau P.,   Volonteri M.,  2005, \mn@doi [The
  Astrophysical Journal] {10.1086/428492}, 623, 23–30

\bibitem[\protect\citeauthoryear{Sesana, Volonteri  \& Haardt}{Sesana
  et~al.}{2007}]{Sesana_2007}
Sesana A.,  Volonteri M.,   Haardt F.,  2007, \mn@doi [Monthly Notices of the
  Royal Astronomical Society] {10.1111/j.1365-2966.2007.11734.x}, 377,
  1711–1716

\bibitem[\protect\citeauthoryear{Sesana, Vecchio  \& Colacino}{Sesana
  et~al.}{2008}]{Sesana_2008_PTA}
Sesana A.,  Vecchio A.,   Colacino C.~N.,  2008, \mn@doi [Monthly Notices of
  the Royal Astronomical Society] {10.1111/j.1365-2966.2008.13682.x}, 390, 192

\bibitem[\protect\citeauthoryear{Sesana, Shankar, Bernardi  \& Sheth}{Sesana
  et~al.}{2016}]{Sesana_2016}
Sesana A.,  Shankar F.,  Bernardi M.,   Sheth R.~K.,  2016, \mn@doi [Monthly
  Notices of the Royal Astronomical Society: Letters] {10.1093/mnrasl/slw139},
  463, L6

\bibitem[\protect\citeauthoryear{{Shang}, {Bryan}  \& {Haiman}}{{Shang}
  et~al.}{2010}]{Shang2010MNRAS.402.1249S}
{Shang} C.,  {Bryan} G.~L.,   {Haiman} Z.,  2010, \mn@doi [\mnras]
  {10.1111/j.1365-2966.2009.15960.x}, \href
  {https://ui.adsabs.harvard.edu/abs/2010MNRAS.402.1249S} {402, 1249}

\bibitem[\protect\citeauthoryear{Shannon et~al.,}{Shannon
  et~al.}{2015}]{Shannon_2015}
Shannon R.~M.,  et~al., 2015, \mn@doi [Science] {10.1126/science.aab1910}, 349,
  1522

\bibitem[\protect\citeauthoryear{{Sijacki}, {Springel}  \&
  {Haehnelt}}{{Sijacki} et~al.}{2009}]{Sijacki2009}
{Sijacki} D.,  {Springel} V.,   {Haehnelt} M.~G.,  2009, \mn@doi [\mnras]
  {10.1111/j.1365-2966.2009.15452.x}, \href
  {https://ui.adsabs.harvard.edu/abs/2009MNRAS.400..100S} {400, 100}

\bibitem[\protect\citeauthoryear{Silk}{Silk}{2017}]{Silk_2017}
Silk J.,  2017, \mn@doi [The Astrophysical Journal] {10.3847/2041-8213/aa67da},
  839, L13

\bibitem[\protect\citeauthoryear{Silk \& Rees}{Silk \&
  Rees}{1998}]{Silk:1997xw}
Silk J.,  Rees M.~J.,  1998, Astron. Astrophys., 331, L1

\bibitem[\protect\citeauthoryear{{S{\k{a}}dowski} \&
  {Gaspari}}{{S{\k{a}}dowski} \& {Gaspari}}{2017}]{Sadowski2017}
{S{\k{a}}dowski} A.,  {Gaspari} M.,  2017, \mn@doi [\mnras]
  {10.1093/mnras/stx543}, \href
  {https://ui.adsabs.harvard.edu/abs/2017MNRAS.468.1398S} {468, 1398}

\bibitem[\protect\citeauthoryear{Smith \& Caldwell}{Smith \&
  Caldwell}{2019}]{Smith_2019}
Smith T.~L.,  Caldwell R.~R.,  2019, \mn@doi [Physical Review D]
  {10.1103/physrevd.100.104055}, 100

\bibitem[\protect\citeauthoryear{{Springel}}{{Springel}}{2005}]{Springel2005Gadget}
{Springel} V.,  2005, \mn@doi [\mnras] {10.1111/j.1365-2966.2005.09655.x},
  \href {https://ui.adsabs.harvard.edu/abs/2005MNRAS.364.1105S} {364, 1105}

\bibitem[\protect\citeauthoryear{Springel \& Hernquist}{Springel \&
  Hernquist}{2003}]{Springel_2003}
Springel V.,  Hernquist L.,  2003, \mn@doi [Monthly Notices of the Royal
  Astronomical Society] {10.1046/j.1365-8711.2003.06206.x}, 339, 289–311

\bibitem[\protect\citeauthoryear{{Springel}, {Di Matteo}  \&
  {Hernquist}}{{Springel} et~al.}{2005}]{Springel2005Modelling}
{Springel} V.,  {Di Matteo} T.,   {Hernquist} L.,  2005, \mn@doi [\mnras]
  {10.1111/j.1365-2966.2005.09238.x}, \href
  {https://ui.adsabs.harvard.edu/abs/2005MNRAS.361..776S} {361, 776}

\bibitem[\protect\citeauthoryear{{Steinborn}, {Dolag}, {Hirschmann}, {Prieto}
  \& {Remus}}{{Steinborn} et~al.}{2015}]{Steinborn2015}
{Steinborn} L.~K.,  {Dolag} K.,  {Hirschmann} M.,  {Prieto} M.~A.,   {Remus}
  R.-S.,  2015, \mn@doi [\mnras] {10.1093/mnras/stv072}, \href
  {https://ui.adsabs.harvard.edu/abs/2015MNRAS.448.1504S} {448, 1504}

\bibitem[\protect\citeauthoryear{Tanaka \& Haiman}{Tanaka \&
  Haiman}{2009}]{Tanaka_2009}
Tanaka T.,  Haiman Z.,  2009, \mn@doi [The Astrophysical Journal]
  {10.1088/0004-637x/696/2/1798}, 696, 1798

\bibitem[\protect\citeauthoryear{Tornatore, Borgani, Dolag  \&
  Matteucci}{Tornatore et~al.}{2007}]{Tornatore_2007}
Tornatore L.,  Borgani S.,  Dolag K.,   Matteucci F.,  2007, \mn@doi [Monthly
  Notices of the Royal Astronomical Society]
  {10.1111/j.1365-2966.2007.12070.x}, 382, 1050–1072

\bibitem[\protect\citeauthoryear{{Valentini}, {Murante}, {Borgani}, {Monaco},
  {Bressan}  \& {Beck}}{{Valentini} et~al.}{2017}]{Valentini2017}
{Valentini} M.,  {Murante} G.,  {Borgani} S.,  {Monaco} P.,  {Bressan} A.,
  {Beck} A.~M.,  2017, \mn@doi [\mnras] {10.1093/mnras/stx1352}, \href
  {https://ui.adsabs.harvard.edu/abs/2017MNRAS.470.3167V} {470, 3167}

\bibitem[\protect\citeauthoryear{{Valentini}, {Borgani}, {Bressan}, {Murante},
  {Tornatore}  \& {Monaco}}{{Valentini} et~al.}{2019}]{Valentini2019}
{Valentini} M.,  {Borgani} S.,  {Bressan} A.,  {Murante} G.,  {Tornatore} L.,
  {Monaco} P.,  2019, \mn@doi [\mnras] {10.1093/mnras/stz492}, \href
  {https://ui.adsabs.harvard.edu/abs/2019MNRAS.485.1384V} {485, 1384}

\bibitem[\protect\citeauthoryear{{Valentini} et~al.,}{{Valentini}
  et~al.}{2020}]{Valentini2020}
{Valentini} M.,  et~al., 2020, \mn@doi [\mnras] {10.1093/mnras/stz3131}, \href
  {https://ui.adsabs.harvard.edu/abs/2020MNRAS.491.2779V} {491, 2779}

\bibitem[\protect\citeauthoryear{{Valentini}, {Gallerani}  \&
  {Ferrara}}{{Valentini} et~al.}{2021}]{Valentini:2021}
{Valentini} M.,  {Gallerani} S.,   {Ferrara} A.,  2021, \mn@doi [\mnras]
  {10.1093/mnras/stab1992}, \href
  {https://ui.adsabs.harvard.edu/abs/2021MNRAS.tmp.1894V} {}

\bibitem[\protect\citeauthoryear{Valiante et~al.,}{Valiante
  et~al.}{2020}]{Valiante_2020}
Valiante R.,  et~al., 2020, \mn@doi [Monthly Notices of the Royal Astronomical
  Society] {10.1093/mnras/staa3395}, 500, 4095–4109

\bibitem[\protect\citeauthoryear{{Vito}, {Di Mascia}, {Gallerani}, {Zana},
  {Ferrara}, {Carniani}  \& {Gilli}}{{Vito} et~al.}{2022}]{Vito2022}
{Vito} F.,  {Di Mascia} F.,  {Gallerani} S.,  {Zana} T.,  {Ferrara} A.,
  {Carniani} S.,   {Gilli} R.,  2022, \mn@doi [\mnras]
  {10.1093/mnras/stac1422}, \href
  {https://ui.adsabs.harvard.edu/abs/2022MNRAS.514.1672V} {514, 1672}

\bibitem[\protect\citeauthoryear{{Vogelsberger} et~al.,}{{Vogelsberger}
  et~al.}{2014}]{Vogelsberger2014Natur}
{Vogelsberger} M.,  et~al., 2014, \mn@doi [\nat] {10.1038/nature13316}, \href
  {https://ui.adsabs.harvard.edu/abs/2014Natur.509..177V} {509, 177}

\bibitem[\protect\citeauthoryear{Volonteri \& Bellovary}{Volonteri \&
  Bellovary}{2012}]{Volonteri_2012}
Volonteri M.,  Bellovary J.,  2012, \mn@doi [Reports on Progress in Physics]
  {10.1088/0034-4885/75/12/124901}, 75, 124901

\bibitem[\protect\citeauthoryear{Volonteri, Haardt  \& Madau}{Volonteri
  et~al.}{2003}]{Volonteri_2003}
Volonteri M.,  Haardt F.,   Madau P.,  2003, \mn@doi [The Astrophysical
  Journal] {10.1086/344675}, 582, 559–573

\bibitem[\protect\citeauthoryear{Volonteri, Reines, Atek, Stark  \&
  Trebitsch}{Volonteri et~al.}{2017}]{Volonteri_2017}
Volonteri M.,  Reines A.~E.,  Atek H.,  Stark D.~P.,   Trebitsch M.,  2017,
  \mn@doi [The Astrophysical Journal] {10.3847/1538-4357/aa93f1}, 849, 155

\bibitem[\protect\citeauthoryear{{Volonteri} et~al.,}{{Volonteri}
  et~al.}{2020}]{Volonteri2020}
{Volonteri} M.,  et~al., 2020, \mn@doi [\mnras] {10.1093/mnras/staa2384}, \href
  {https://ui.adsabs.harvard.edu/abs/2020MNRAS.498.2219V} {498, 2219}

\bibitem[\protect\citeauthoryear{{Weinberger} et~al.,}{{Weinberger}
  et~al.}{2017}]{Weinberger2017}
{Weinberger} R.,  et~al., 2017, \mn@doi [\mnras] {10.1093/mnras/stw2944}, \href
  {https://ui.adsabs.harvard.edu/abs/2017MNRAS.465.3291W} {465, 3291}

\bibitem[\protect\citeauthoryear{{White} \& {Rees}}{{White} \&
  {Rees}}{1978}]{White_Rees}
{White} S.~D.~M.,  {Rees} M.~J.,  1978, \mn@doi [\mnras]
  {10.1093/mnras/183.3.341}, \href
  {https://ui.adsabs.harvard.edu/abs/1978MNRAS.183..341W} {183, 341}

\bibitem[\protect\citeauthoryear{{Wiersma}, {Schaye}  \& {Smith}}{{Wiersma}
  et~al.}{2009}]{Wiersma2009MNRAS}
{Wiersma} R. P.~C.,  {Schaye} J.,   {Smith} B.~D.,  2009, \mn@doi [\mnras]
  {10.1111/j.1365-2966.2008.14191.x}, \href
  {https://ui.adsabs.harvard.edu/abs/2009MNRAS.393...99W} {393, 99}

\bibitem[\protect\citeauthoryear{Woods et~al.,}{Woods
  et~al.}{2019}]{Woods_2019}
Woods T.~E.,  et~al., 2019, \mn@doi [Publications of the Astronomical Society
  of Australia] {10.1017/pasa.2019.14}, 36

\bibitem[\protect\citeauthoryear{Wyithe \& Loeb}{Wyithe \&
  Loeb}{2003}]{Wyithe_2003}
Wyithe J. S.~B.,  Loeb A.,  2003, \mn@doi [The Astrophysical Journal]
  {10.1086/375187}, 590, 691–706

\bibitem[\protect\citeauthoryear{Yoshida, Omukai  \& Hernquist}{Yoshida
  et~al.}{2008}]{Yoshida_2008}
Yoshida N.,  Omukai K.,   Hernquist L.,  2008, \mn@doi [Science]
  {10.1126/science.1160259}, 321, 669

\bibitem[\protect\citeauthoryear{Yue, Ferrara, Salvaterra, Xu  \& Chen}{Yue
  et~al.}{2013}]{Yue_2013}
Yue B.,  Ferrara A.,  Salvaterra R.,  Xu Y.,   Chen X.,  2013, \mn@doi [Monthly
  Notices of the Royal Astronomical Society] {10.1093/mnras/stt826}, 433,
  1556–1566

\bibitem[\protect\citeauthoryear{{Zana}, {Gallerani}, {Carniani}, {Vito},
  {Ferrara}, {Lupi}, {Di Mascia}  \& {Barai}}{{Zana} et~al.}{2022}]{Zana2022}
{Zana} T.,  {Gallerani} S.,  {Carniani} S.,  {Vito} F.,  {Ferrara} A.,  {Lupi}
  A.,  {Di Mascia} F.,   {Barai} P.,  2022, \mn@doi [\mnras]
  {10.1093/mnras/stac978}, \href
  {https://ui.adsabs.harvard.edu/abs/2022MNRAS.tmp..961Z} {}

\bibitem[\protect\citeauthoryear{eLISA Consortium et~al.,}{eLISA Consortium
  et~al.}{2013}]{consortium2013gravitational}
eLISA Consortium T.,  et~al., 2013, The Gravitational Universe (\mn@eprint
  {arXiv} {1305.5720})

\makeatother
\end{thebibliography}
